\documentclass[a4paper,10pt,oneside,final]{article}
\pdfoutput=1

\usepackage{siunitx}
\usepackage{cite}
\usepackage{amsmath}
\usepackage{booktabs}
\usepackage{xcolor}
\usepackage{mathtools}
\mathtoolsset{showonlyrefs}

\newcommand  {\R}{\mathbf{R}}
\newcommand  {\1}{\mathbf{1}}

\newcommand  {\NIG}{\text{NIG}}

\newcommand  {\N}{\text{N}}
\renewcommand{\S}{\text{S}}
\newcommand  {\pdf}{\text{pdf}}

\renewcommand{\Re}{\text{Re}}

\newcommand  {\sign}{\operatorname{sign}}
\newcommand  {\vol}{\operatorname{vol}}
\newcommand  {\<}{\langle}
\renewcommand{\>}{\rangle}

\title{A causal continuous-time stochastic model for the turbulent
  energy cascade in a helium jet flow}

\author{Emil Hedevang (emil@imf.au.dk)\\
  Department of Mathematics, Aarhus University,\\
  Ny Munkegade 118, 8000 Aarhus C, Denmark\\
  \\
  J\"urgen Schmiegel (schmiegl@imf.au.dk)\\
  Department of Engineering, Aarhus University,\\
  Ny Munkegade 118, 8000 Aarhus C, Denmark
}

\begin{document}
\maketitle

\begin{abstract}
  \noindent
  We discuss continuous cascade models and their potential for
  modelling the energy dissipation in a turbulent flow. Continuous
  cascade processes, expressed in terms of stochastic integrals with
  respect to L\'evy bases, are examples of ambit processes. These
  models are known to reproduce experimentally observed properties of
  turbulence: The scaling and self-scaling of the correlators of the
  energy dissipation and of the moments of the coarse-grained energy
  dissipation.  We compare three models: a normal model, a normal
  inverse Gaussian model and a stable model. We show that the normal
  inverse Gaussian model is superior to both, the normal and the
  stable model, in terms of reproducing the distribution of the energy
  dissipation; and that the normal inverse Gaussian model is superior
  to the normal model and competitive with the stable model in terms
  of reproducing the self-scaling exponents. Furthermore, we show that
  the presented analysis is parsimonious in the sense that the
  self-scaling exponents are predicted from the one-point distribution
  of the energy dissipation, and that the shape of these distributions
  is independent of the Reynolds number.
\end{abstract}


\section{Introduction}

Since the pioneering work of Kolmogorov~\cite{Ko-1962} and
Obou\-khov~\cite{Ob-1962}, where the turbulent energy dissipation is
assumed to be log-normally distributed, the small-scale intermittency
of the energy dissipation in turbulence has received much
attention~\cite{Fr-1995,Sr-An-1997}. The small scale intermittency is
primarily expressed in terms of multifractal and universal scaling of
inertial range statistics, including extended
self-similarity~\cite{Be-et-al-1993}, scaling and self-scaling of
correlators~\cite{Sc-2005}, and the statistics of breakdown
coefficients~\cite{Cl-Sc-Gr-2008}.

Early attemps to model the rapid variation of the turbulent velocity
field include \cite{Vi-Ba-1991, Be-et-al-1993b, El-et-al-1995,
  Ju-et-al-1994, Bi-et-al-1998, Ar-Ba-Mu-1998} (among many
others). Such phenomenological approaches are sometimes called
``synthetic turbulence'' and can be divided into two classes. The
first direction starts from modelling the velocity field and derives
the model for the energy dissipation by taking squared small scale
increments. The second line of investigation focuses on modelling the
energy dissipation field and derives the velocity field by various,
partly ad hoc, manipulations. The approach presented here focuses on
modelling the energy dissipation as the fundamental field, not derived
from an a priori velocity field.

In~\cite{Vi-Ba-1991}, an iterative, geometric multi-affine model for the
one-dimensional velocity process is contructed and some of the basic,
global statistical quantities of the energy dissipation field are
derived. However, this discrete, dyadic approach does not allow to
give explicit expressions for more specific statistical quantities. In
the approach discussed here, all main statistical quantities, like
$n$-point correlations and probability densities can be derived
analytically since our approach defines the energy dissipation field
as an explicit closed expression that is mathematically tractable and
does not involve an iterative procedure.

Another dyadic, iterative approach for the construction of the
velocity field is discussed in~\cite{Be-et-al-1993b}. Their model is
based on a wavelet decomposition of the velocity field combined with a
multiplicative cascading structure for the wavelet coefficients. The
energy dissipation field is then derived from small scale increments,
again leaving only limited room for an exact analytical treatment of
higher order statistics. However, as discussed
in~\cite{El-et-al-1995}, such wavelet approaches are superior over
discrete geometric approaches as they allow to model stationarity in a
mathematical more rigorous way. The approach discussed here does not
suffer from problems related to mathematical rigour and no iterative
limit arguments are needed for the construction. A related and
interesting wavelet-based approach is discussed
in~\cite{Bi-et-al-1998}, which allows for a sequential construction of
the field. A further wavelet-based approach~\cite{Ar-Ba-Mu-1998}
builds on random functions and their orthogonal wavelet fransform. The
authors show that to each such random function there is an associated
cascade on a dyadic tree of wavelet coefficients. The performance of
the model is illustrated by numerical examples, with little analytical
insight.

The models~\cite{Vi-Ba-1991, Be-et-al-1993b} fail to incorporate
skewness for the velocity increments~\cite{Ju-et-al-1994}, a basic
property of turbulent fields. As an alternative approach,
\cite{Ju-et-al-1994}~proposes a combination of a multiplicative
cascade for the energy dissipation, the use of Kolmogorov's refined
similarity hypothesis~\cite{Ko-1962} and an appropriate summation rule
for the increments to construct the velocity field. Here, again, only
discrete iterative procedures are employed which make analytical
statistical statements very difficult.

It is important to mention that the continuous cascade fields
discussed in this paper are potentially useful in the above cited
works \cite{Be-et-al-1993b, El-et-al-1995, Ju-et-al-1994,
  Bi-et-al-1998, Ar-Ba-Mu-1998} as a candidate for a closed continuous
and mathematically tractable and flexible version of the discrete
multiplicative procedures employed there.

Discrete and continuous random cascade processes have proved useful in
describing phenomenologically the small-scale behaviour of the
turbulent energy dissipation~\cite{Ma-1974, Fr-Su-Ne-1978,
  Be-Pa-Pa-Vu-1984, Sc-Lo-1987, Me-Sr-1991, Jo-Li-Gr-1999, Cl-Gr-2000,
  Jo-Gr-Li-2000}. In~\cite{Cl-et-al-2005} the surrogate energy
dissipation is modelled as a discrete random multiplicative cascade
process. Choosing the law of the cascade generators to be log-normal
yields the Kolmogorov-Oboukhov model. A continuous analogue to the
discrete multiplicative cascade processes is formulated in terms of
integrals with respect to L\'evy bases and has been shown
\cite{BaNi-Sc-2004, Sc-et-al-2004, Sc-2005} to be computationally
tractable and to accurately describe the two- and three-point
statistics of the energy dissipation.

In the cited works, focus is on the modelling of $n$-point statistics
of the energy dissipation, not the distribution of the energy
dissipation itself. Indeed, \cite{Sc-2005} concludes with a remark
that the law of the L\'evy basis driving the cascade model should be
inferred and its dependency on the Reynolds number should be
investigated. 

Both, discrete and continuous multiplicative cascade processes,
suggest that the law of the logarithm of the energy dissipation should
be infinitely divisible. Infinite divisibility is necessary for the
cascade models discussed here to be defined in a mathematical
rigorous way. Furthermore, it greatly simplifies analytic calculations
of some of the statistical properties of the models. Among the
infinitely divisible distributions are the normal, stable, and normal
inverse Gaussian distributions.  These three classes of distributions
each have their own tail behaviour.

The use of stable L\'evy bases for modelling of the energy dissipation
has been investigated in~\cite{Cl-Sc-Gr-2008} and it is concluded by
analysing the breakdown coefficients that ``except for the log-normal
limit, this leaves no room for the log-stable modelling of the
turbulent energy cascade.'' The present paper investigates the
alternative of using a normal inverse Gaussian L\'evy basis to model
the energy dissipation and addresses the one-point distributions,
multifractality of the coarse-grained energy dissipation, and
two-point statistics.

The use of normal inverse Gaussian distributions in turbulence
modelling is not new. In~\cite{Cl-et-al-2005}, the parameters of the
normal inverse Gaussian distribution cascade generator are estimated
from scaling exponents and cumulants, which are moment estimates,
notorious for their sensitivity to outliers. Indeed, estimation of the
normal inverse Gaussian parameters (and those of other distributions)
may not be feasible from sample moments. In this paper we will apply
likelihood methods which do not suffer from the same problems as the
moment based methods.

Our motivation for the use of the normal inverse Gaussian distribution
is not based on physical arguments. We mainly exploit that these
distributions are flexible, yet simple, and capable of attaining a
wide range of shapes. Some shortcomings of the normal inverse Gaussian
distribution will be discussed. Being a normal mean-variance mixture
with an inverse Gaussian as mixing distribution, the tail behavior of
the mixture is related to that of the mixing
distribution~\cite{BaNi-Ke-Soe-1982}. Hence one obtains a recipe for
the construction of distributions with a prescribed tail
behaviour. The infinite divisibility is necessary for the calculus of
the stochastic integrals used in this paper.

The She-Leveque-Dubrulle model~\cite{Sh-Le-1994,Du-1994,Sh-Wa-1995},
which has been shown to accurately predict the scaling exponents of the
coarse-grained energy dissipation and the structure functions of the
velocity increments, prescibes a Poisson distribution for the
log-energy dissipation. We find that the Poisson distribution provides
a poor fit and that the normal inverse Gaussian distribution is a
clear improvement. It should be noted that, being infinitely
divisible, the L\'evy-It\^o decomposition expresses the normal inverse
Gaussian distribution as a linear combination of Poisson
distributions.

%
%
The paper is organised as follows. Section~\ref{sec:data} provides
some background on the data analysed in this paper.
Section~\ref{sec:model} recalls the construction of continuous cascade
processes in terms of integrals with respect to L\'evy bases.
Section~\ref{sec:helium} applies the theory to the data and shows how
the distribution of the surrogate energy dissipation determines the
scaling and self-scaling exponents of the two-point correlators of the
energy dissipation and the coarse-grained energy dissipation.
Section~\ref{sec:conc} concludes. The two appendices provide necessary
background on the normal inverse Gaussian distribution and integration
with respect to L\'evy bases.


\section{Background on the data}
\label{sec:data}

%
%
We analyse thirteen data sets, each consisting of approximately
sixteen million one-point time records of the velocity component in
the mean stream direction in helium gas jet
flow~\cite{Ch-et-al-2000}. The time series can be assumed to be
stationary. In \cite{Cl-et-al-2004}, eleven of the thirteen data sets
are used in an analysis of the intermittency exponent of the turbulent
energy cascade. In particular, \cite[table I]{Cl-et-al-2004}
summarises useful information about the data sets. Data sets no.~6
and~9 are not considered in~\cite{Cl-et-al-2004}.

%
%
Let~$u$ denote the velocity component in the mean stream direction,
and let~$U$ denote the mean stream velocity. Since the flow can be
assumed to be homogeneous and isotropic, we use the surrogate energy
dissipation $\epsilon(x) = 15\nu(\partial u/\partial x)^2$ as a proxy
for the energy dissipation and henceforth omit the ``surrogate''
predicate. Here~$x$ denotes the position along the mean stream
direction, and~$\nu$ denotes the viscosity. We apply the Taylor frozen
flow hypothesis to express the energy dissipation in terms of the
measured time series.  We do not invoke an ``instantaneous Taylor
correction''~\cite{Ch-et-al-2000} since it introduces spurious effects
in correlators of the energy dissipation. Since any change of the
energy dissipation by a multiplicative constant is inconsequential for
the conclusions, we scale the energy dissipation to have unit
mean. Finally, the derivative $\partial u/\partial t$ is calculated
from the discrete samples using interpolation with third-order
splines.

The resolution scale $\Delta x$ (mean velocity $U$ times sampling time
$\Delta t$) is $1$--$5$ times the Kolmogorov length~$\eta =
(\nu^3/\langle\epsilon\rangle)^{1/4}$, and the Taylor-microscale based
Reynolds number~$\Re_\lambda$ varies from~$85$ to~$1181$.
Table~\ref{tab:summary} lists~$\Re_\lambda$ in addition to parameters
that will be explained later.

%
%
\begin{figure}
  \centering
  \includegraphics{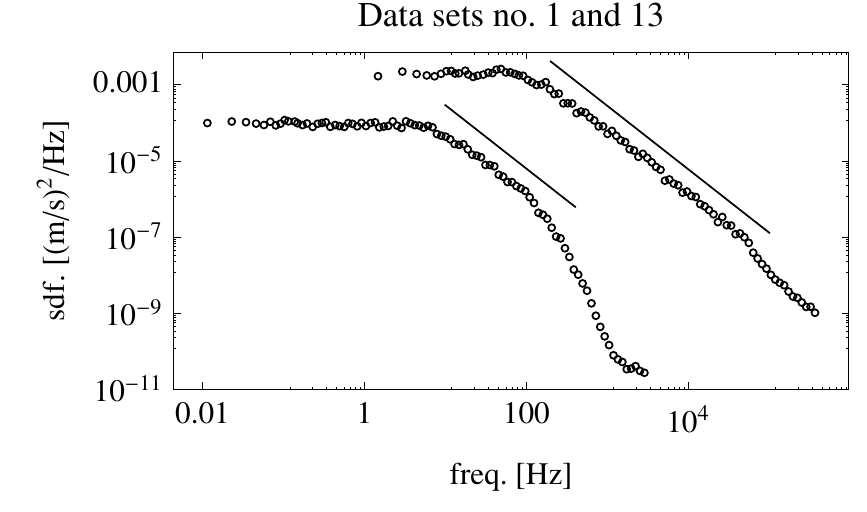}
  \caption{The spectral density function for data set no.~1 (lower)
    and no.~13 (upper).  The slope of the lines is $-5/3$.}
  \label{fig:sdf}
\end{figure}
\begin{figure}
  \centering
  \includegraphics{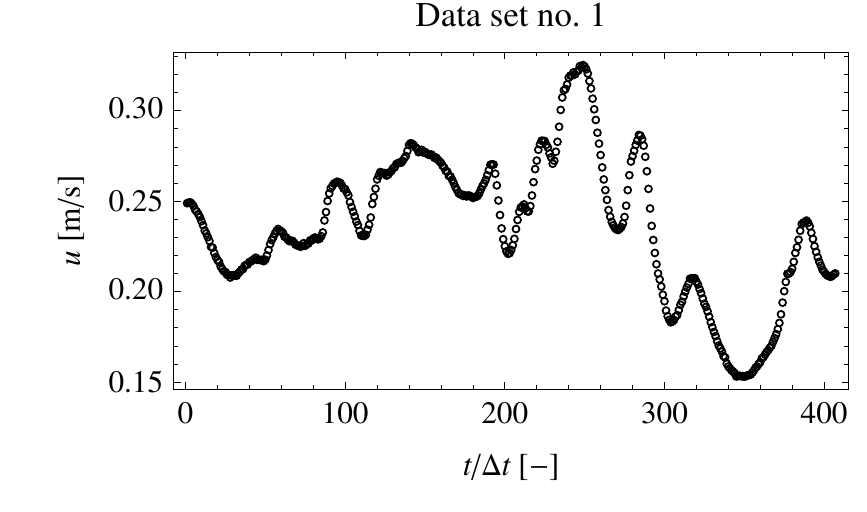}\\
  \includegraphics{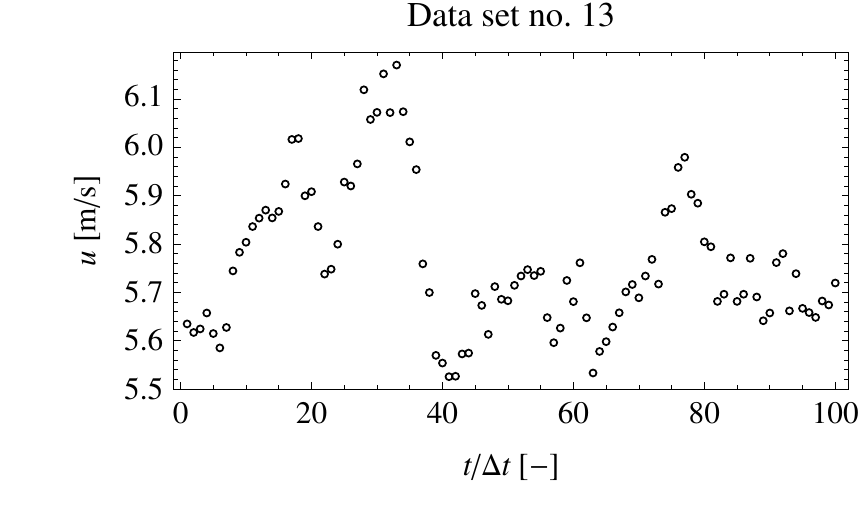}
  \caption{Exerpts of the time series for data sets no.~1 and~13. The
    sampling resolution of data set no.~1 is approximately four times
    that of data set no.~13. The length of both exerpts is
    approximately $500\eta$.}
  \label{fig:ts}
\end{figure}
The spectral density of the velocity component in the mean stream
direction is estimated using Welsh's overlapping segment averages with
a Hanning taper, a block length of $500\,000$, and an overlap
of~$50\%$, see e.g.~\cite{Pe-Wa-1993} for details about the method.
Figure~\ref{fig:sdf} shows the spectral density for data sets no.~1
and~13. The inertial ranges are found at frequencies from
approximately \SI{10}{Hz} to \SI{100}{Hz} and \SI{200}{Hz} to
\SI{60}{kHz}, respectively, after which the transition to the
dissipation range occurs.  The slope in the inertial range is
approximately $-5/3$, obeying Kolmogorov's $5/3$-law.  For most of the
data sets, the spectral density attains an almost constant value at
the highest frequencies. We interpret this as instrument noise. As
shown by fig.~\ref{fig:ts} (top) this does not appear to distort the
data significantly. For some of the data sets, the resolution scale is
almost five times the Kolmogorov length. The effect of this becomes
apparent in fig.~\ref{fig:ts} (bottom) where the time series appears
much less smooth than in fig.~\ref{fig:ts} (top) where the resolution
scale is close to the Kolmogorov length. If data set no.~1 is
subsampled by retaining every fourth record, a picture similar to
fig.~\ref{fig:ts}~(bottom) is obtained.  Having no means to improve
the resolution scale of the measured data sets, we consider
downsampling the data to improve the smoothness so that calculation of
the derivative $\partial u/\partial t$ becomes reliable. As will be
remarked later, the influence of the downsampling on the shape of the
one-point distribution of the energy dissipation is negligible and it
significanly decreases the scatter of all quantities derived from
moments estimates. Henceforth all graphs and tables refer to data that
has been downsampled by a factor of two. We found little or no change
when downsampling by a factor of three or four compared to a factor of two.

%
%
\begin{figure*}
  \centering
  \includegraphics{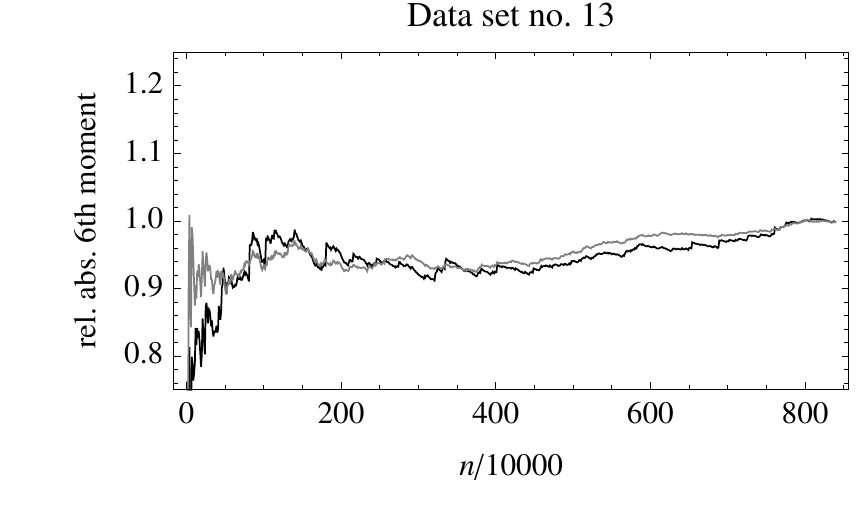}\\
  \includegraphics{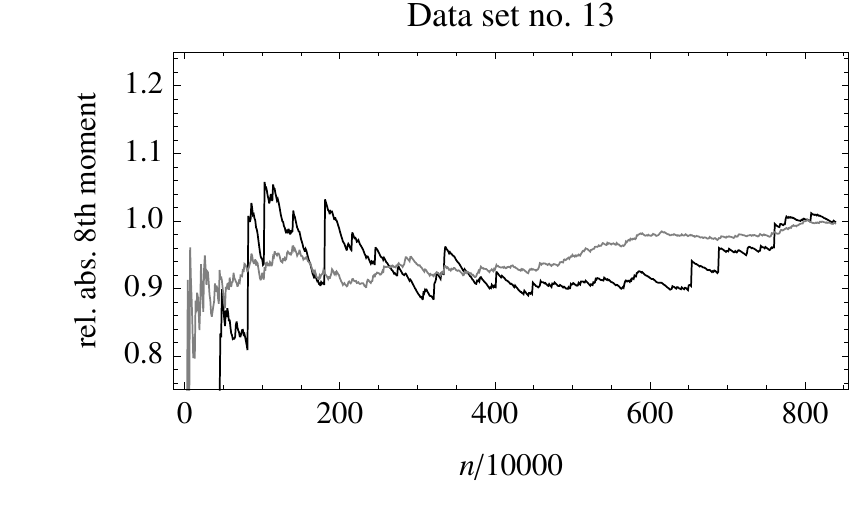}
  \caption{Relative absolute sample moment of order 6 (top) and 8
    (bottom) of $\partial u/\partial t$ before (black) and after (gray)
    removal of outliers for data set no.~13.}
  \label{fig:mom}
\end{figure*}
To assess the existence of moments, we consider, for a given time
series $y=\{y_1,\ldots,y_N\}$, the relative absolute sample moment
$M^p_n(y)/M^p_N(Y)$ of order~$p$ and sample size~$n\leq N$, where
$M^p_n(y)= \frac1n \sum_{i=1}^n |y_i|^p$. Figure~\ref{fig:mom} shows
the relative absolute sample moments of $\partial u/\partial t$ of
orders~6 and~8 for data set no.~7. We see that the sample moments are
corrupted by outliers and that removal of outliers increases the
reliability of the moment estimates. We will assume that $\partial
u/\partial t$ has moments of order at least~8. The removal of outliers
does not distort the parameters of the one-point distribution of the
energy dissipation.


\section{A L\'evy based continuous multiplicative cascade model}
\label{sec:model}

In this section we present the model for the energy dissipation. The
model provides a link between the distribution and the two-point
correlators of the energy dissipation.

%
%
In~\cite{Sc-2005,Sc-BaNi-Eg-2005} the energy dissipation~$\epsilon$ is
modelled as a $(1+1)$-dimensional stochastic process (one dimension in
space and one in time) given as the exponential of an integral with
respect to a L\'evy basis~$Z$ on $\R^2$,
\begin{align}
  \epsilon(x,t) 
  = \exp\Bigl(\int_{A(x,t)} Z(dx'\,dt')\Bigr) = \exp(Z(A(x,t))),
\end{align}
where $A(x,t)\subseteq\R^2$ is called the \emph{ambit set}. We will
assume that the L\'evy basis is homogeneous (see
appendix~\ref{app:levy}), and that the ambit sets are defined by
$A(x,t) =(x,t)+A$ for some bounded set $A\subseteq\R^2$. Thus we
ensure that~$\epsilon$ is stationary in space and time. The
model~\eqref{eq:epsilon} is an example of a random multiplicative
cascade process in continuous space and time. Since only one-point
time series are available for our analysis, we will ignore the spatial
dependency and consider the energy dissipation as a function of time
alone and at a fixed position in space,
\begin{align}
  \label{eq:epsilon}
  \epsilon(t) 
  = \exp\Bigl(\int_{A(t)} Z(dx'\,dt')\Bigr) = \exp(Z(A(t))),
\end{align}
where $A(t) = A + (0,t)$. However, it must be emphasised that the
model is equally capable of modelling specific spatial statistics in
one or more spatial dimensions.

The details of the derivation of various properties
of~\eqref{eq:epsilon} are given in~\cite{Sc-2005,Sc-BaNi-Eg-2005}. For
convenience we repeat in the following three subsections the formulas
we need for the subsequent analysis.

\subsection{One-point distributions}

The distribution of $\log\epsilon$ is given by~\eqref{eq:kumulant}
or~\eqref{eq:cumulant}. Since the integrand in~\eqref{eq:epsilon} is
constant, we have that
\begin{align}
  \label{eq:klogepsilon}
  K(s\ddagger\log\epsilon(t)) = K(s\ddagger Z')\vol(A),
\end{align}
where $K(s\ddagger X) = \log\<\exp(sX)\>$ denotes the logarithm of the
Laplace transform of the random variable~$X$, $Z'$ denotes the L\'evy
seed corresponding to the L\'evy basis~$Z$, and $\vol(A)$ denotes the
volume (or area) of the set~$A$.
By~\eqref{eq:klogepsilon}, the distribution of the energy dissipation
is closely related to the distribution of the L\'evy seed~$Z'$. For
brevity we define $K(s) = K(s\ddagger Z')$ and $K(p,q) =
K(p+q)-K(p)-K(q)$.

\subsection{Coarse-grained energy dissipation}

Let $\epsilon_l$ denote the coarse-grained energy dissipation,
\begin{align}
  \epsilon_l(t) = \frac1l\int_{t-l/2}^{t+l/2}\epsilon(s)\,ds.
\end{align}
Kolmogorov's refined similarity hypothesis states that
\begin{align}
  u(t+l) - u(t) \stackrel{\text{law}}{=} V(l\epsilon_l)^{1/3},
\end{align}
where $V$ is a random variable independent of the energy
dissipation. This hypothesis provides a relation between the scaling
exponents $\zeta_p$ of the moments $\<(u(t+l) - u(t))^p\>\sim
l^{\zeta_p}$ of longitudinal velocity differences and the scaling
exponents $\sigma_p$ of the moments $\<\epsilon_l^p\>\sim
l^{\sigma_p}$ of the coarse-grained energy dissipation, namely
\begin{align}
  \zeta_p = p/3 + \sigma_{p/3}.
\end{align}
The exponent $\sigma_2$ is the intermittency exponent which quantifies
the deviation from Kolmogorov's 1941 theory where $\zeta_p = p/3$.

From the scaling property $\<\epsilon_l^p\>\sim l^{\sigma_p}$ we
derive the self-scaling property $\<\epsilon_l^q\> \sim
\<\epsilon_l^p\>^{\sigma_q/\sigma_p}$ where the ratio $\sigma_q/\sigma_p$ is the
self-scaling exponent.  It is shown in~\cite{Sc-BaNi-Eg-2005}
that~\eqref{eq:epsilon} implies that
\begin{align}
  \sigma_p \approx C_A(K(p) - pK(1))
\end{align}
for integral values of $p$ where $C_A$ is a constant depending only on
the size and shape of the ambit set~$A$. It follows that the
self-scaling exponent $\sigma_q/\sigma_p$ is independent of the ambit
set and hence only depends on the distribution of the L\'evy
seed~$Z'$.

\subsection{Two-point correlators}

%
%
The two-point correlator $c_{p,q}$ of order $(p,q)$ of the energy
dissipation is defined as
\begin{align}
  c_{p,q}(l) =
  \frac{\langle\epsilon(l)^p\epsilon(0)^q\rangle}
  {\langle\epsilon(l)^p\rangle\langle\epsilon(0)^q\rangle}.
\end{align}
The sum $p+q$ is the \emph{total order} of the correlator $c_{p,q}$.
It is shown in~\cite{Sc-2005,Sc-BaNi-Eg-2005} that
\begin{align}
  \label{eq:cepsilon}
  \begin{aligned}
    c_{p,q}(l) &= \exp\bigl(K(p,q)\vol(A(l)\cap
    A(0))\bigr).
  \end{aligned}
\end{align}
We observe that the exponent in~\eqref{eq:cepsilon} is expressed as a
product where the first factor depends only on the L\'evy basis and
the order of the correlator, and the second factor depends only on the
overlap of the ambit sets. This provides a way of modelling a wide
range of correlators, since the shape of the ambit set, under suitable
assumptions, can be determined from the correlator, see~\cite{Sc-2005}
for details or subsec.~\ref{subsec:ambit-set} for a particular
example.

%
%
Experiments~\cite{Cl-et-al-2004,Sc-et-al-2004} reveal that, at least
for $p+q\leq 3$, the two-point correlators exhibit \emph{scaling},
\begin{align}
  \label{eq:s}
  c_{p,q}(l) 
  &\propto l^{-\tau(p,q)},
\end{align}
in a range of $l$ comparable to the inertial range of the velocity
structure functions. It is straightforward to show that the scaling
exponent~$\tau(1,1)$ is equal to the intermittency
exponent~$\sigma_2$. This follows from the fact that the second
derivative of $\<l^2\epsilon_l^2(t)\>$ with respect to~$l$ behaves as
$c_{1,1}(l)$~\cite{Sc-BaNi-Eg-2005}.

In~\cite{Sc-2005} it is shown that the two-point correlators also
enjoy the property of \emph{self-scaling},
\begin{align}
  \label{eq:ss}
  c_{p_2,q_2}(l)
  = c_{p_1,q_1}(l)^{\tau(p_1,q_1;p_2,q_2)},
\end{align}
for an even wider range of $l$, in analogy to extended
self-similarity~\cite{Be-et-al-1993} of structure functions. Here
$\tau(p_1,q_1;p_2,q_2) = \tau(p_2,q_2)/\tau(p_1,q_1)$ is the
self-scaling exponent.

%
%

By~\eqref{eq:cepsilon} we immediately obtain self-scaling of the
correlators under the model~\eqref{eq:epsilon},
\begin{align}
  c_{p_2,q_2}(l)
  = c_{p_1,q_1}(l)^{K(p_1,q_1;p_2,q_2)},
\end{align}
where
\begin{align}
  K(p_1,q_1;p_2,q_2) = K(p_2,q_2)/K(p_1,q_1)
\end{align}
is the self-scaling exponent. As noted in~\cite{Sc-2005}, the
self-scaling property is independent of the shape of the ambit set and
thus scaling of the correlators is not necessary for self-scaling of
the correlators. The L\'evy seed $Z'$ determines
through~\eqref{eq:klogepsilon} the distribution of the energy
dissipation and therefore the self-scaling exponents.  Both, the
distribution and the self-scaling exponents, can be estimated from
data and hence compared with the model.

%
%
The model~\eqref{eq:epsilon} is also able to reproduce empirical
three-point correlators~\cite{Sc-et-al-2004}, but these are not
considered in the present paper.

%
%
The correlators are moment estimates. Therefore they may not exist
beyond a certain order. When they exist, they are sensitive to noise
and outliers, particularly at high orders. By H\"older's inequality we
have
\begin{align}
  \<\epsilon(t_1)^p\epsilon(t_2)^q\> 
  \leq \<\epsilon(t_1)^{p+q}\>^{p/(p+q)}
  \<\epsilon(t_2)^{p+q}\>^{q/(p+q)}.
\end{align}
It follows that the correlator $c_{p,q}$ exists provided the velocity
derivative $\partial u/\partial t$ has finite moment of order
$2(p+q)$. Under the model~\eqref{eq:epsilon} it follows
by~\eqref{eq:klogepsilon} that $c_{p,q}$ exists if and only if the
L\'evy seed $Z'$ has exponential moments of order $p+q$.

\subsection{Ambit sets and scaling of correlators}
\label{subsec:ambit-set}

\begin{figure}
  \centering
  \includegraphics{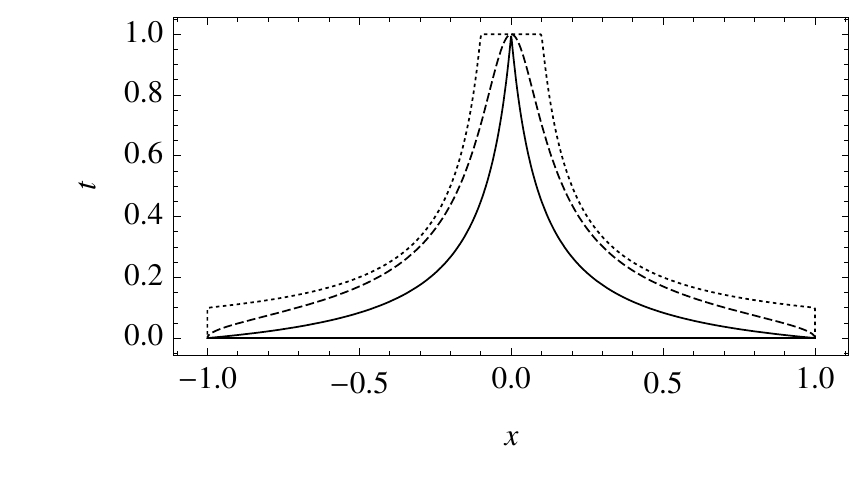}
  \caption{The boundary of the ambit set~\protect\eqref{eq:A} for the
    values $\theta=1$ (solid), $\theta=2$ (long dashes), and
    $\theta=100$ (short dashes). The other parameters are $T=1$ and
    $L=10$.}
  \label{fig:ambit-set}
\end{figure}
While the self-scaling of the correlators and the self-scaling of the
coarse-grained energy dissipation is independent of the shape of the
ambit set~$A$, the scaling property~\eqref{eq:s} requires~$A$ to be
specified appropriately according to~\eqref{eq:cepsilon}. In this
subsection we consider one such specification. We define for $T>0$,
$L>1$, and $\theta>0$ the function~$g$ by
\begin{align}
  \label{eq:g}
  g(t) = \Bigl( \frac{1-(t/T)^\theta}{1+(t/(T/L))^\theta}
  \Bigr)^{1/\theta} \quad 0\leq t\leq T.
\end{align}
In the limit $\theta=\infty$ we have that
\begin{align}
  g(t) =
  \begin{cases}
    1 & 0\leq t \leq T/L, \\
    T/(Lt) & T/L < t \leq T.
  \end{cases}
\end{align}
The ambit set $A$ is given as
\begin{align}
  \label{eq:A}
  A = \{(x,t)\mid 0\leq t\leq T, |x| \leq g(t)\}.
\end{align}
Figure~\ref{fig:ambit-set} shows examples of the ambit set for three
values of the parameter~$\theta$. By~\eqref{eq:cepsilon},
\begin{align}
  \label{eq:cg}
  c_{p,q}(l) 
  &= \exp\Bigl(K(p,q)\int_l^T 2g(s)\,ds\Bigr).
\end{align}
It follows that
\begin{align}
  \label{eq:s-epsilon}
  \frac{d\log c_{p,q}(l)}{d\log l}
  \approx -K(p,q)2T/L,
  \quad T/L\ll l\ll T.
\end{align}
Hence $K(p,q)2T/L$ is the scaling exponent of $c_{p,q}$. In the
limit $\theta=\infty$ we have perfect scaling, 
\begin{align}
  \frac{d\log c_{p,q}(l)}{d\log l}
  =-K(p,q)2T/L,
  \quad T/L<l<T.
\end{align}
The parameter $\theta$ is merely a tuning parameters to account for
imperfect scaling of the correlators. In view of~\eqref{eq:cepsilon},
where the overlap of the ambit sets determines the correlator, $T$ may
be interpreted as the decorrelation time, and $T/L$ is a time scale
near and below which the scaling behaviour has terminated.

\subsection{Three L\'evy bases}
\label{subsec:levy-bases}

%
%
We consider three distributions of the L\'evy seed~$Z'$: 
\begin{align}
  Z'\sim
  \begin{cases}
    \N(\mu,\delta) 
    &\text{normal (N)}, \\
    \S(\alpha,\beta,\mu,\delta)
    &\text{stable (S)}, \\
    \NIG(\alpha,\beta,\mu,\delta) 
    &\text{normal inv. Gaussian (NIG)}.
  \end{cases}
\end{align}
The normal L\'evy seed yields a log-normal model for the energy
dissipation, consistent with the Kol\-mo\-go\-rov-Oboukhov
theory~\cite{Ob-1962}. The stable L\'evy seed is included to enable
comparison with~\cite{Cl-Sc-Gr-2008}. The use of a normal inverse
Gaussian L\'evy seed can be motivated in several ways, though none of
them is based on physical arguments. The normal inverse Gaussian
distribution is computationally tractable and allows a wide range of
distributions (see fig.~\ref{fig:nig-tri}).  Furthermore, the
representation~\eqref{eq:NIGmix} shows that the normal inverse
Gaussian distributions form a natural generalisation of the normal
distributions by allowing the variance of the normal distribution to
be random. Since the variance is positive, and if there is a ``typical
variance'', it is natural to model the variance using an unimodal
distribution on the positive real line. A flexible distribution of
this type is the inverse Gaussian law as shown in
fig.~\ref{fig:ig-pdf}.

While other infinitely distributions could be considered as well, the
three classes of distributions above exhaust a wide range of unimodal
distributions with a specific tail behaviour: The normal distributions
have ``light tails'', the stable distributions have ``heavy tails'',
decaying algebraically, and the normal inverse Gaussian distributions
have ``semi-heavy tails'', decaying exponentially,
see~\eqref{eq:semi-heavy}.

The common symbols for the parameters $(\alpha,\beta,\mu,\delta)$ have
been chosen due to the similarity of their interpretations. In all
three cases,~$\mu$ is a location parameter, $\delta$ is a scale
parameter, $\alpha$ and~$\beta$ are shape parameters, and~$\beta$
determines the asymmetry of the distribution. The domain of the
parameters is $\mu\in\R$ and $\delta>0$ in all three cases;
$0<\alpha\leq2$ and $-1\leq\beta\leq1$ in the stable case; and
$|\beta|<\alpha$ in the normal inverse Gaussian case. The
parametrisation of the stable distribution is chosen to
follow~\cite[eq.~(1.1.6)]{Sa-Ta-1994}, so that the log-characteristic
function is given by
\begin{align}
  &\log \<\exp(is\S(\alpha,\beta,\mu,\delta))\>\\
  &\quad= \begin{cases}
    i\mu s-\delta^\alpha|s|^\alpha(1-i\beta\sign(s)\tan(\pi\alpha/2)),
    & \alpha\neq1,\\
    i\mu s-\delta|s|(1-i\beta\frac2\pi\sign(s)\log|s|),
    & \alpha = 1.
  \end{cases}
\end{align}

%
%
It follows from~\eqref{eq:kumulant} or~\eqref{eq:cumulant} that the
distribution of $\log\epsilon$ in each of the three cases is
\begin{align}
  \label{eq:logepsilondist}
  \log\epsilon \sim
  \begin{cases}
    \N(\vol(A)\mu,\vol(A)\delta)
    &Z'\sim\N,\\
    \S(\alpha,\beta,\vol(A)\mu,\vol(A)^{1/\alpha}\delta)
    &Z'\sim\S,\\
    \NIG(\alpha,\beta,\vol(A)\mu,\vol(A)\delta)
    &Z'\sim\NIG.
  \end{cases}
\end{align}
We see that the shape parameters of the distribution of $\log\epsilon$
are, in all three cases, identical to the shape parameters of the
distribution of the L\'evy seed~$Z'$, and that the location and scale
parameters of $\log\epsilon$ are multiplied with a factor determined
by the size of the ambit set.

%
%
The normal distribution has exponential moments of all orders.
Provided $\beta=-1$, the stable distribution has exponential moments
of all non-negative orders. We will therefore assume that $\beta=-1$
in the stable case. The normal inverse Gaussian distribution has
exponential moments of order~$s$ if 
\begin{align}
  \label{eq:nigcond}
  s<\alpha-\beta.
\end{align}
In these cases the log-Laplace transform of the L\'evy seed is given
by
\begin{align}
  \label{eq:KZ'}
  K(s)
  =\begin{cases}
    \mu s + \frac12\delta s^2 
    &Z'\sim\N, \\
    \mu s -\bigl(\delta^\alpha/\cos(\pi\alpha/2)\bigr)s^\alpha
    &S'\sim\S,\\
    \mu s +\delta\bigl(\sqrt{\alpha^2-\beta^2}-\sqrt{\alpha^2-(\beta+s)^2}\bigr)
    & Z'\sim\NIG.
  \end{cases}
\end{align}
It follows that
\begin{align}
  \label{eq:Kpq2}
  K(p,q) = \kappa(p,q)\cdot
  \begin{cases}
    \frac12\delta
    &Z'\sim\N,\\
    -\bigl(\delta^\alpha/\cos(\pi\alpha/2)\bigr)
    &Z'\sim\S,\\
    \delta
    &Z'\sim\NIG,
  \end{cases}
\end{align}
where
\begin{align}
  \label{eq:kappa}
  \kappa(p,q) =
  \begin{cases}
    pq 
    &Z'\sim\N,\\
    (p+q)^\alpha - p^\alpha - q^\alpha
    &Z'\sim\S,\\[2pt]
    \begin{aligned}
      &\sqrt{\alpha^2-(\beta+p)^2}+\sqrt{\alpha^2-(\beta+q)^2}\\ 
      &\quad -\sqrt{\alpha^2-\beta^2}-\sqrt{\alpha^2-(\beta+p+q)^2}\\ 
    \end{aligned}
    &Z'\sim\NIG.
  \end{cases}
\end{align}
Note that under the normal inverse Gaussian model, we must require
that $p+q<\alpha-\beta$ to ensure the existence of the correlator of
total order $p+q$.  Note also that~$\kappa$ depends only on the order
$(p,q)$ and the shape parameters. The self-scaling exponents of the
correlators are now given by
\begin{align}
  \label{eq:kappass}
  K(p_1,q_1;p_2,q_2)
  = \frac{\kappa(p_2,q_2)}{\kappa(p_1,q_1)}.
\end{align}
In a similar manner it follows that the self-scaling exponents
$\sigma_q/\sigma_p$ of the coarse-grained energy dissipation depend
only on the order $(p,q)$ and the shape parameters of the L\'evy seed.
We therefore conclude that both kinds of self-scaling exponents are
uniquely determined by the orders under consideration and the shape of
the one-point distribution of the energy dissipation.


\section{Data analysis}
\label{sec:helium}

In this section we apply the model from section~\ref{sec:model} to the
thirteen data sets. The one-point distribution of the energy
dissipation is used to predict the scaling and self-scaling exponents
of the two-point correlators and the coarse-grained energy
dissipation.

\subsection{One-point distributions}

%
%
\begin{figure}
  \centering
  \includegraphics{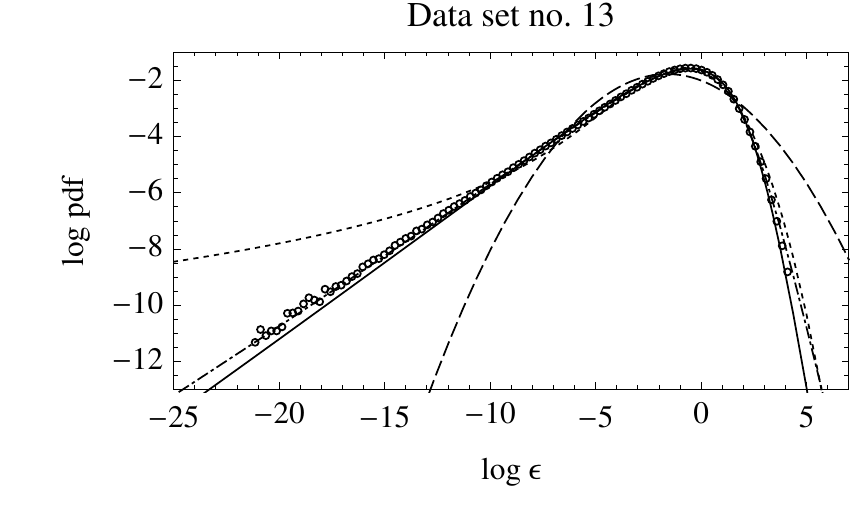}
  \caption{Distribution of the logarithm of the energy dissipation
    represented by the logarithm of the probability density function
    for data set no.~13 downsampled by a factor two: data (circles),
    normal fit (long dashes), stable fit (short dashes), normal
    inverse Gaussian fit (dots and dashes), and normal inverse
    Gaussian fit constrained to possess finite exponential moments of
    order $4.5$ (black curve).}
  \label{fig:pdf}
\end{figure}
Figure~\ref{fig:pdf} shows that the distribution of $\log\epsilon$ has
a distinct, non-Gaussian, asymmetric shape. The shape is independent
of the downsamling applied to smooth the data. Four parametric
distributions are fitted to the data using likelihood methods: normal,
stable, normal inverse Gaussian, and normal inverse Gaussian
constrained to have finite exponential moment of order~$4.5$ according
to~\eqref{eq:nigcond}. Clearly, the normal inverse Gaussian
distribution provides an excellent fit for all amplitudes. The left
tail is overestimated by the stable distribution. The unconstrained
normal inverse Gaussian distribution is also able to accurately fit
the distribution, though it slightly underestimates the left tail. In
what follows, any reference to the normal inverse Gaussian
distribution will imply the constrained version.

It is necessary to constrain the shape parameters of the normal
inverse Gaussian distribution to ensure the existence of the
correlators up to a certain total order. This constrainment is a
delicate issue.  Firstly, it introduces the problem of determining an
appropriate upper order. And secondly, it implies a slightly poorer
fit. The choice of $4.5$ as the upper order is somewhat arbitrary but
provides a compromise between fitting the distribution of the energy
dissipation accurately and predicting the self-scaling exponents of
the correlators and the coarse-grained energy dissipation. We
conjecture the existence of a more appropropriate infinitely divisible
distribution, perhaps within the class of normal mean-variance
mixtures. For the present paper, the normal inverse Gaussian will
suffice.

%
%
\begin{table}
  \centering\small
  \hbox to\hsize{\hss\tabcolsep=4.5pt
    \begin{tabular}[c]{cccccccccccc}
    &&\multicolumn{4}{c}{%
      $\overbrace{\hphantom{\hskip34mm}}^{\textstyle\text{normal inverse Gaussian}}$}
    &\multicolumn{3}{c}{%
      $\overbrace{\hphantom{\hskip24mm}}^{\textstyle\text{stable}}$}\\
    & $\Re_\lambda$ 
    & $\alpha$ & $\beta$ 
    & $\mu_\epsilon$ & $\delta_\epsilon$ 
    & $\alpha$ 
    & $\mu_\epsilon$ & $\delta_\epsilon$ 
    & $\Delta x/\eta$
    & $\tau(1,1)$ & $\tau'(1,1)$
    \\
    \midrule
    $1$ & $85$ & $2.49$ & $-2.01$ & $2.24$ & $2.77$ & $1.62$ & $-1.81$
    & $1.31$ & $2.39$ & $0.111$ & $0.120$ \\
    $2$ & $89$ & $2.49$ & $-2.01$ & $2.30$ & $2.87$ & $1.63$ & $-1.83$
    & $1.32$ & $2.09$ & $0.139$ & $0.128$ \\
    $3$ & $124$ & $2.50$ & $-2.00$ & $2.35$ & $2.96$ & $1.63$ &
    $-1.84$ & $1.33$ & $1.94$ & $0.114$ & $0.102$ \\
    $4$ & $208$ & $2.51$ & $-1.99$ & $2.60$ & $3.34$ & $1.67$ &
    $-1.97$ & $1.42$ & $3.50$ & $0.146$ & $0.154$ \\
    $5$ & $209$ & $2.50$ & $-2.00$ & $2.31$ & $2.91$ & $1.63$ &
    $-1.83$ & $1.32$ & $1.92$ & $0.112$ & $0.083$ \\
    $6$ & $283$ & $2.50$ & $-2.00$ & $2.40$ & $3.06$ & $1.65$ &
    $-1.86$ & $1.35$ & $2.53$ & $0.110$ & --- \\
    $7$ & $352$ & $2.51$ & $-1.99$ & $2.57$ & $3.32$ & $1.67$ &
    $-1.95$ & $1.40$ & $4.49$ & $0.129$ & $0.130$ \\ 
    $8$ & $463$ & $2.51$ & $-1.99$ & $2.54$ & $3.31$ & $1.67$ &
    $-1.92$ & $1.39$ & $3.76$ & $0.128$ & $0.092$ \\ 
    $9$ & $703$ & $2.50$ & $-2.00$ & $2.38$ & $3.03$ & $1.65$ &
    $-1.86$ & $1.34$ & $4.91$ & $0.108$ & --- \\ 
    $10$ & $885$ & $2.50$ & $-2.00$ & $2.45$ & $3.13$ & $1.65$ &
    $-1.89$ & $1.37$ & $6.72$ & $0.107$ & $0.089$ \\ 
    $11$ & $929$ & $2.50$ & $-2.00$ & $2.44$ & $3.11$ & $1.65$ &
    $-1.88$ & $1.36$ & $7.16$ & $0.112$ & $0.079$ \\ 
    $12$ & $985$ & $2.50$ & $-2.00$ & $2.43$ & $3.07$ & $1.65$ &
    $-1.89$ & $1.36$ & $9.53$ & $0.104$ & $0.105$ \\ 
    $13$ & $1181$ & $2.49$ & $-2.01$ & $2.30$ & $2.88$ & $1.63$ &
    $-1.83$ & $1.33$ & $9.72$ & $0.097$ & $0.061$ \\ 
  \end{tabular}\hss}
  \caption{Summary of the thirteen data sets. The Taylor micro-scale
    Reynolds numbers are from~\cite{Ch-et-al-2000}. The normal inverse
    Gaussian parameters (constrained to yield finite exponential moment of
    order~$4.5$ according to~\protect\eqref{eq:nigcond}) and the stable
    parameters are estimated from the
    distribution~\protect\eqref{eq:logepsilondist} of $\log\epsilon$ using
    likelihood methods. $\Delta x$ denotes in units of the Kolmogorov
    length~$\eta$ the effective spatial resolution after downsampling by a factor
    of two. The effect of downsampling by a factor of two changes the
    normal inverse Gaussian shape parameters $\alpha$ and $\beta$ by
    around \SI{0.4}\% and the stable shape parameter $\alpha$ by around
    \SI{2}\%. $\tau(1,1)$ is the intermittency exponent estimated by
    fitting~\protect\eqref{eq:g} and~\protect\eqref{eq:cg} to the
    correlator~$c_{1,1}$, and $\tau'(1,1)$ is the intermittency exponent
    determined in~\protect\cite{Cl-et-al-2004}.}
  \label{tab:summary}
\end{table}

Table~\ref{tab:summary} summarises the estimated parameters for each
of the thirteen data sets. Note that for both, the normal inverse
Gaussian distribution and the stable distribution, the estimated shape
parameters do not depend on the Taylor micro-scale Reynolds number,
while the location and scale parameters vary only slightly. This is a
clear indication of universality of the distribution of the energy
dissipation, at least within the thirteen data sets considered
here. Downsampling the data by a factor of two changes the normal
inverse Gaussian shape parameters $\alpha$ and $\beta$ by only around
\SI{0.4}\% and the stable shape parameter~$\alpha$ by around \SI{2}\%.

\subsection{Coarse-grained energy dissipation}

\begin{figure*}
  \centering
  \includegraphics{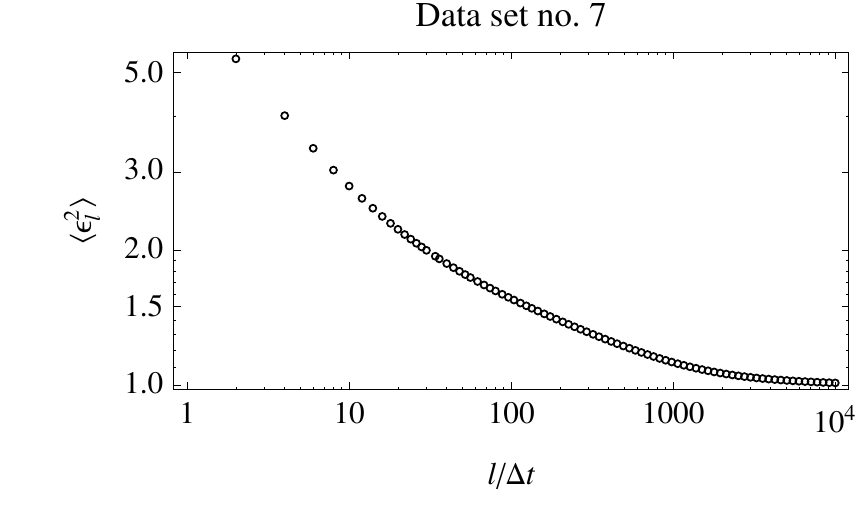}
  \includegraphics{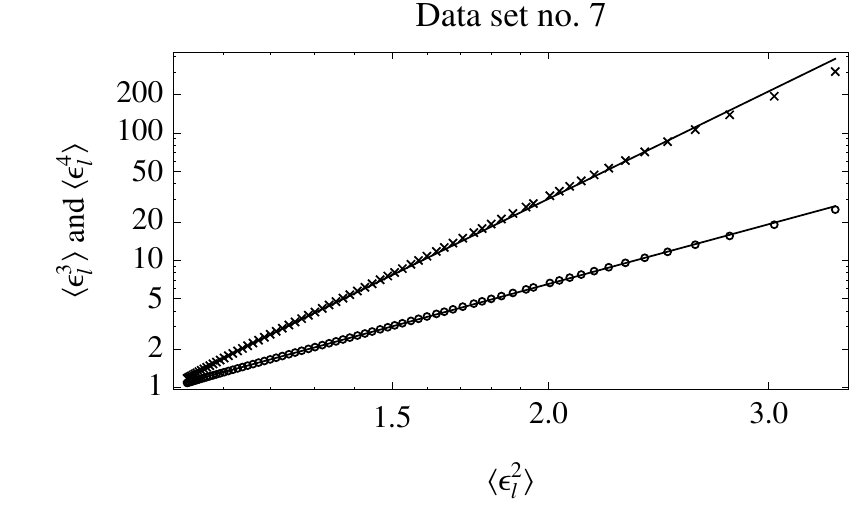}
  \caption{Scaling and self-scaling of the moments $\<\epsilon_l^p\>$
    of the coarse-grained energy dissipation for data set no.~7. Top:
    $\<\epsilon_l^2\>$ as a function of the lag~$l$. The scaling
    behaviour is far from pronounced. Bottom: $\<\epsilon_l^p\>$ for
    $p=3,4$ as a function of $\<\epsilon_l^2\>$. The self-scaling
    behaviour is clear.}
  \label{fig:cgs}
\end{figure*}

\begin{figure*}
  \centering
  \includegraphics{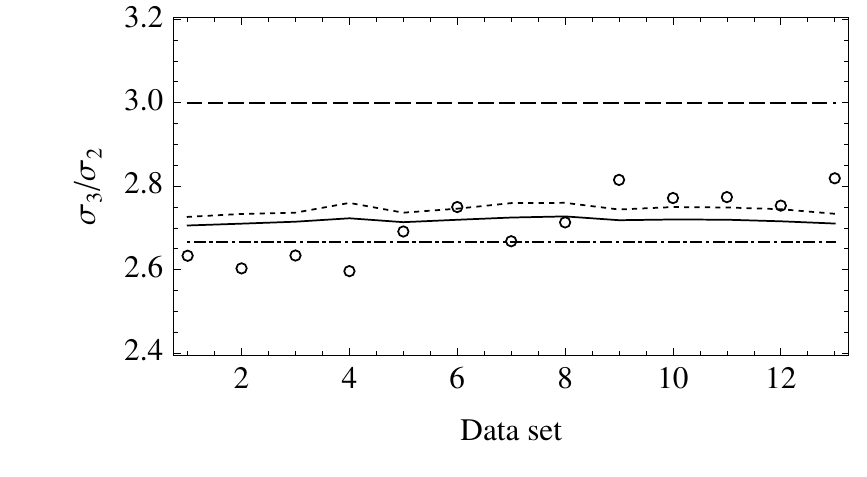}
  \includegraphics{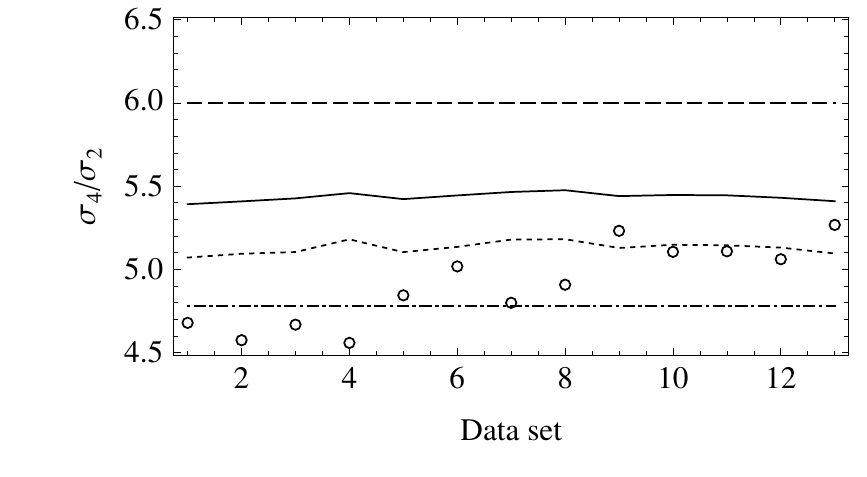}
  \caption{Self-scaling exponents $\sigma_3/\sigma_2$ (top) and
    $\sigma_4/\sigma_2$ (bottom) of the moments $\<\epsilon_l^p\>$ of
    the coarse-grained energy dissipation: data (circles), normal
    model (long dashes), stable model (short dashes), normal inverse
    Gaussian model (solid), and She-Leveque-Dubrulle model (dots and
    dashes).}
  \label{fig:cgsse}
\end{figure*}

The behaviour of the moments $\<\epsilon_l^p\>$ as a function of the
lag~$l$ does not reveal any pronounced scaling range, see
fig.~\ref{fig:cgs} (top) as an example. Compare with fig.~\ref{fig:s}
which demonstrates a clear scaling range for the two-point
correlators. However, as shown by fig.~\ref{fig:cgs} (bottom) the
moments display a clear self-scaling behaviour, and the self-scaling
exponents~$\sigma_q/\sigma_p$ can be extracted.

The estimated self-scaling exponents can now be compared with the
predictions of the model~\eqref{eq:epsilon} and with the
She-Leveque-Dubrulle model. The former predicts
\begin{align}
  \frac{\sigma_q}{\sigma_p} 
  = \frac{K(q) - qK(1)}{K(p) - pK(1)},
\end{align}
while the latter predicts
\begin{align}
  \frac{\sigma_q}{\sigma_p} =\frac{-2q/3 + 2(1+(2/3)^q)}{-2p/3 +
    2(1+(2/3)^p)}.
\end{align}
Figure~\ref{fig:cgsse} (top) shows that the normal inverse Gaussian
and the stable model agree well with both data and the
She-Leveque-Dubrulle model for the self-scaling exponent
$\sigma_3/\sigma_2$. For $\sigma_4/\sigma_2$ fig.~\ref{fig:cgsse}
(bottom) displays a somewhat greater discrepancy between the
models. The She-Leveque-Dubrulle model agrees best with data for the
data sets with the lowest $\Re_\lambda$ while the stable model agrees
best for the other data sets. The normal inverse Gaussian model
slightly overpredicts the self-scaling exponents, but in general it
provides a prediction of comparable accuracy.

The normal inverse Gaussian model can be forced to align better with
the stable model and the She-Leveque-Dubrulle model by requiring
exponential moments of order~$5$ to exist. This, however, compromises
the prediction of the self-scaling exponents of the two-point
correlators (subsec.~\ref{subsec:tpcss}).

\subsection{Two-point correlators: scaling}

%
%
Since the sample moments of the velocity derivative become
increasingly unreliable with increasing order, we consider only the
orders $(p,q)$ for which $p$~and~$q$ are positive half-integers with
$p+q\leq 3$. This leads to~36 non-trivial combinations in the analysis
of self-scaling of the correlators. The self-scaling exponents satisfy
\begin{align}
  \tau(p_1,q_1;p_3,q_3)
  = \tau(p_1,q_1;p_2,q_2) \tau(p_2,q_2;p_3,q_3),
\end{align}
and it is therefore sufficient to consider eight combinations.

%
%
\begin{figure}
  \centering
  \includegraphics{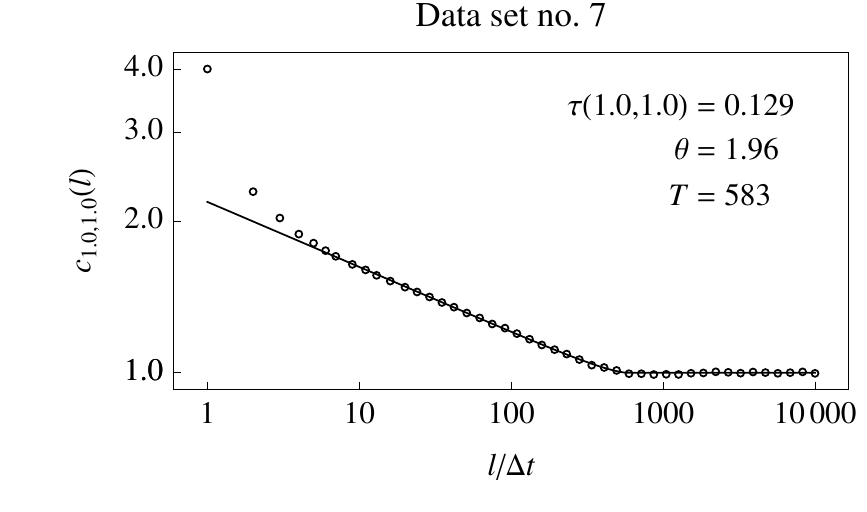}
  \caption{The two-point correlator~$c_{1,1}$ as a function of
    lag~$l$ for data set no.~7 in double logarithmic
    representation (circles). The solid curve shows the least squares
    fit of~\protect\eqref{eq:cg} in the double logarithmic
    representation. The estimated scaling exponent $\tau(1,1)$,
    decorrelation time~$T$, and auxiliary parameter~$\theta$ are shown
    in the figure.}
  \label{fig:s}
\end{figure}

Figure~\ref{fig:s} shows that the correlator~$c_{1,1}$ for data set
no.~7 exhibits scaling for $10 \Delta t\leq l\leq200 \Delta t$. The
figure also shows that the correlator~\eqref{eq:cg} determined
by~\eqref{eq:g} provides a very good fit to the data for lags
$l\geq10 \Delta t$. The parameter $\theta$ determines the behaviour of the
correlator for lags close to the decorrelation time~$T$. At lags $l<10
\Delta t$, the correlators are corrupted by surrogacy
effects~\cite{Cl-Gr-Sr-2003}.  Lags $l<10 \Delta t$ have therefore
been excluded from the fit. Similar results hold for the other data
sets and other orders of the correlators.

The parameter~$L$, which determines the extend of the scaling range of
the correlators, cannot be identified since surrogacy effects corrupt
the correlator at small lags. We have chosen $L=10^4$ as it allows for
clear scaling at all lags $1 \Delta t\leq l\ll T \Delta t$ for all
thirteen data sets.

%
%
The scaling exponents, under model~\eqref{eq:epsilon}, follow from
\eqref{eq:s-epsilon}, \eqref{eq:logepsilondist}, and~\eqref{eq:Kpq2},
\begin{align}
  \label{eq:tau-kappa}
  \tau(p,q) &= C_{L,\theta}\cdot\kappa(p,q)\cdot
  \begin{cases}
    \frac12\delta_\epsilon
    &Z'\sim\N,\\
    \displaystyle\frac{-\delta_\epsilon^\alpha}{\cos(\pi\alpha/2)}
    &Z'\sim\S,\\
    \delta_\epsilon
    &Z'\sim\NIG,
  \end{cases}
\end{align}
where $C_{L,\theta}$ is a factor which depends only on~$L$
and~$\theta$. Since~$L$ cannot be reliably identified by the data
available,~\eqref{eq:tau-kappa} does not allow us to predict the
correlator scaling exponents from the distribution of the energy
dissipation alone.

The scaling exponents may still be estimated directly by
fitting~\eqref{eq:g} and~\eqref{eq:cg} to the
correlators. Table~\ref{tab:summary} shows the estimated intermittency
exponent $\tau(1,1)$ where it is also compared to the estimate found
in~\cite{Cl-et-al-2004}. The two estimates are in reasonable
agreement. The differences are likely due to the differences in the
estimation procedures. We note that for all the data sets, the
intermittency exponent is approximately half of the prediction under
the She-Leveque-Dubrulle model where $\tau(1,1)=2/9$.

%
%
As a final consistency check, we note that, according
to~\eqref{eq:s-epsilon}, the scaling exponents satisfy a relation of
the form
\begin{align}
  \label{eq:tau-check}
  \tau(p,q) = a_{p+q} - a_p - a_q.
\end{align}
By~\eqref{eq:KZ'}, we may choose $a_1=0$. The other unknowns are found
by solving the linear least squares system corresponding
to~\eqref{eq:tau-check}. The root mean square of the relative error
$(a_{p+q} - a_p - a_q)/\tau(p,q)-1$ is below \SI{1}\% except for data
sets no. 1, 2, and 5 where it is \SI{3}\%, \SI{9}\%, and \SI{8}\%,
respectively. We conclude that~\eqref{eq:tau-check} is fulfilled to a
satisfying degree.

\subsection{Two-point correlators: self-scaling}
\label{subsec:tpcss}
%
%
\begin{figure}
  \centering
  \hbox to\textwidth{\hss
    \begin{tabular}{ccc}
      \includegraphics{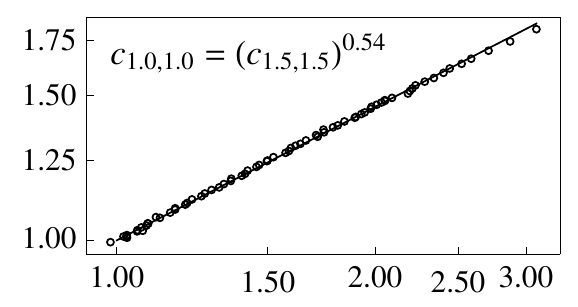} &
      \includegraphics{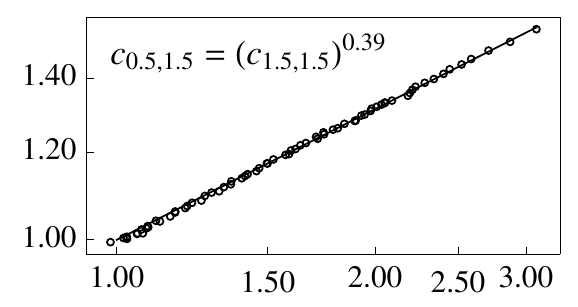} &
      \includegraphics{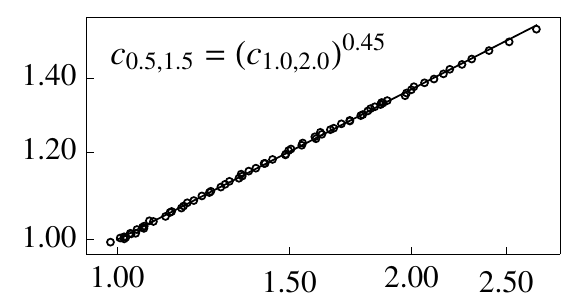} \\
      \includegraphics{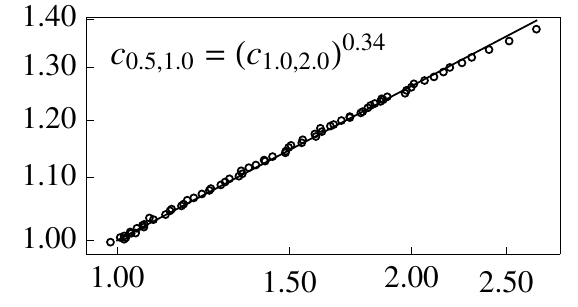} &
      \includegraphics{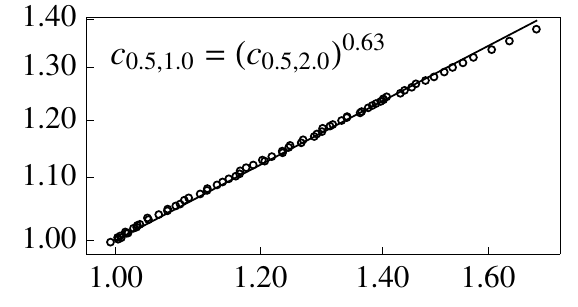} &
      \includegraphics{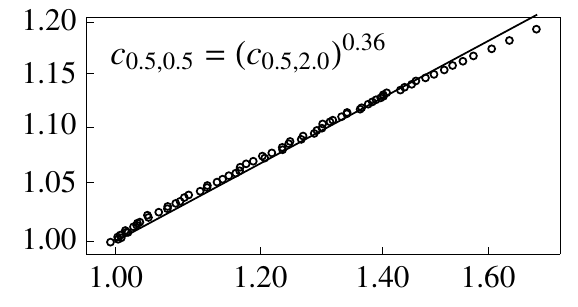} \\
      \includegraphics{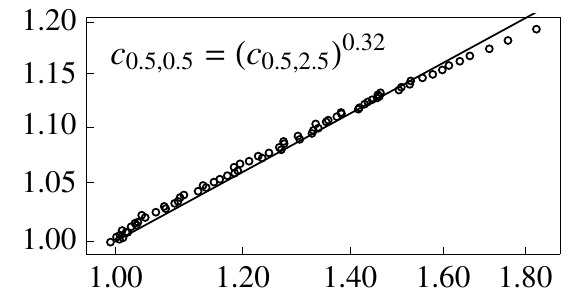} &
      \includegraphics{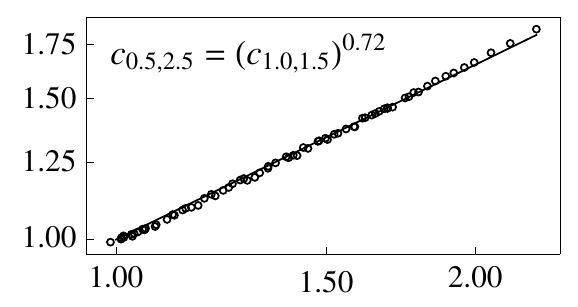} &
      \includegraphics{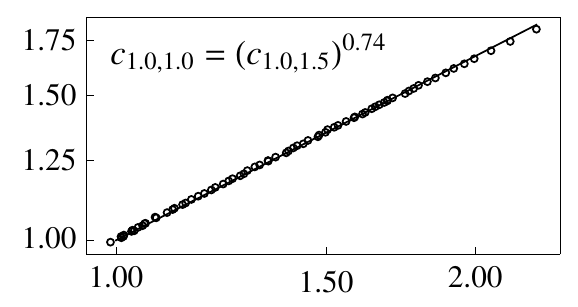} 
    \end{tabular}\hss}
  \caption{The two-point correlator $c_{p_2,q_2}$ ($y$-axis) as a
    function of the two-point correlator $c_{p_1,q_1}$ ($x$-axis) for
    data set no.~7 in double logarithmic representation. The black
    line indicates the self-scaling behaviour $c_{p_2,q_2} =
    (c_{p_1,q_1})^{\tau(p_1,q_1;p_2,q_2)}$ as shown in each subfigure.
    Similar results hold for the other data sets.}
  \label{fig:ss}
\end{figure}
\begin{figure*}
  \centering
  \hbox to\textwidth{\hss
    \begin{tabular}{ccc}
      \includegraphics{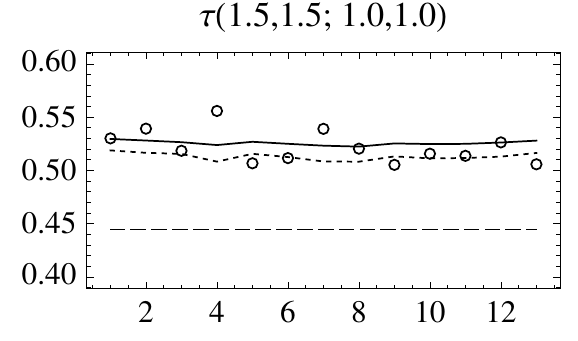} &
      \includegraphics{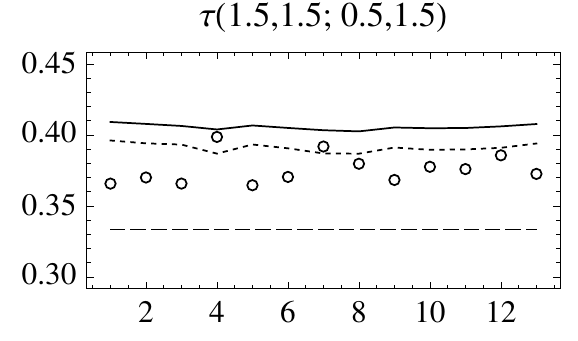} &
      \includegraphics{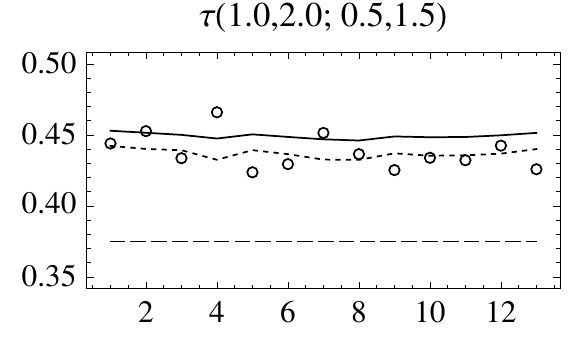} \\
      \includegraphics{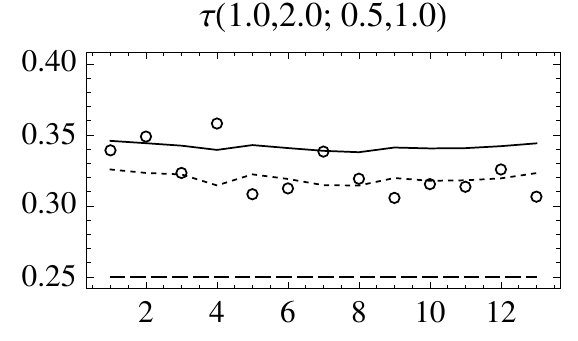} &
      \includegraphics{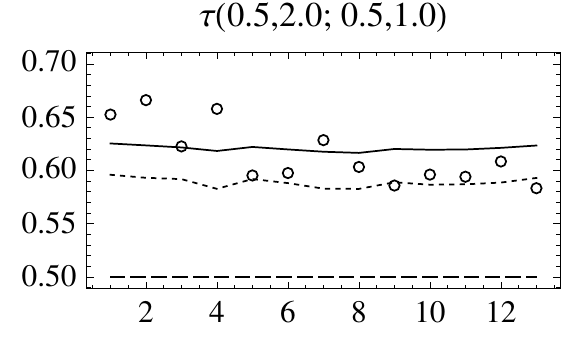} &
      \includegraphics{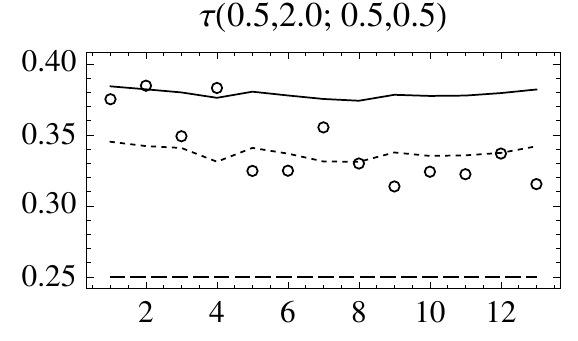} \\
      \includegraphics{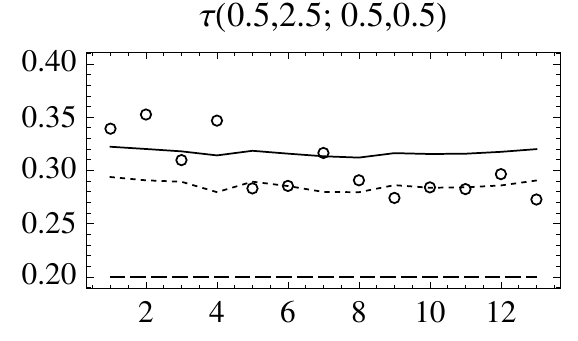} &
      \includegraphics{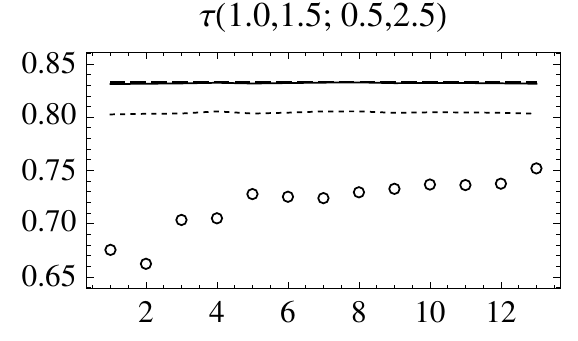} &
      \includegraphics{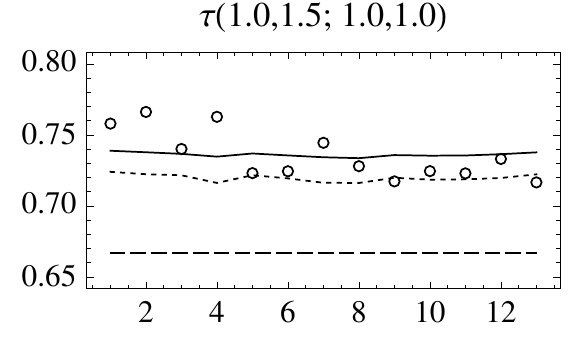} 
    \end{tabular}\hss}
  \caption{The self-scaling exponents~$\tau(p_1,q_1;p_2,q_2)$
    ($y$-axis) for each of the thirteen data sets: data (circles),
    normal model (long dashes), stable model (short dashes), normal
    inverse Gaussian model (solid).}
  \label{fig:sse}
\end{figure*}
The self-scaling of the correlators is confirmed by fig.~\ref{fig:ss}.
Under model~\eqref{eq:epsilon}, the self-scaling exponent
$\tau(p_1,q_2;p_2,q_2)$ is given by the ratio
$\kappa(p_2,q_2)/\kappa(p_1,q_1)$. It depends only on the shape
parameters of the distribution of $\log\epsilon$ and on the order of
the involved correlators. In particular, the self-scaling exponents
are independent of the ambit set and the location and scale parameters
of~$\log\epsilon$. Therefore, the self-scaling exponents are predicted
directly from the estimated shape parameters listed in
tab.~\ref{tab:summary}, and they inherit the
$\Re_\lambda$-independence from $\log\epsilon$.

Figure~\ref{fig:sse} shows the estimated self-scaling exponents for
all the data sets. In most cases the predictions of the normal inverse
Gaussian model and the stable model are within the variation of the
data point, though the normal inverse Gaussian model tends to predict
larger values than the stable model. Both models are clearly superior
to the normal model.  The stable model appears to be slightly more
accurate than the normal inverse Gaussian model.  The case
$(1.0,1.5;0.5,2.5)$ is exceptional as all models fail to predict the
observed self-scaling exponents by approximately $10\%$.

At low Reynolds numbers, the self-scaling exponents tend to deviate
from the self-scaling exponents at higher Reynolds numbers. At low
Reynolds numbers the inertial range is shorter. Whether this affects
the estimation of the self-scaling exponents remains to be
investigated.

%
%
Further observations may be drawn from fig.~\ref{fig:sse}. Firstly,
the predicted self-scaling exponents are in all three cases almost
independent of the Taylor micro-scale Reynolds number. This follows
from the fact that the estimated shape parameters of the distribution
of $\log\epsilon$ are almost independent of $\Re_\lambda$
(tab.~\ref{tab:summary}). As noted previously, this independence hints
at a universal character of the self-scaling exponents and the
distribution of $\log\epsilon$. Secondly, the shape parameters are
derived from one-point statistics, while the self-scaling exponents
are derived from two-point statistics. The predictability of the
latter from the former hints at a parsimonious description of the
energy dissipation as the exponential of an integral with respect to a
L\'evy basis: Higher order statistics are reasonably well predicted
from lower order statistics.


\section{Conclusion and outlook}
\label{sec:conc}

%
%
Thirteen time series of one-point measurements of the velocity
component in the mean stream direction in a helium jet are analysed
from the point of view of the (surrogate) energy dissipation. The
distribution of the logarithm of the energy dissipation is shown to be
well approximated by normal inverse Gaussian distributions, a property
possessed by neither the normal distribution nor the stable
distribution. Furthermore, the shape of the distribution is apparently
universal, i.e., it does not depend on the Reynolds number, at least
within the considered data sets.

The two-point correlators of the energy dissipation show scaling and
self-scaling. By modelling the energy dissipation as the exponential
of an integral with respect to a L\'evy basis, a connection is
established between the shape of the distribution of the logarithm of
the energy dissipation and the scaling and self-scaling exponents of
the two-point correlators.  Thus the model is parsimonious: the
two-point correlators are all, to good accuracy, determined by the
one-point distribution. The normal inverse Gaussian distributions are
compared to the stable distributions. While the stable distributions
are also capable of modelling the scaling and self-scaling exponents,
they are far from capable of accurately modelling the distribution of
the energy dissipation (fig.~\ref{fig:pdf}). In particular, the stable
model implies infinite variance of $\log\epsilon$, yet the data
suggest that $\log\epsilon$ has finite moments at least up to
order~6. The use of normal inverse Gaussian distributions allows
accurate modelling of both the correlators and the distribution of the
energy dissipation.

The model also allows prediction of the self-scaling exponents of the
moments of the coarse-grained energy dissipation. Comparison to the
predictions of the She-Leveque-Dubrulle model is made.

While the normal inverse Gaussian distributions allow fairly accurate
modelling of the one-point distribution, two-point correlators of the
energy dissipation, and the self-scaling exponents of the moments of
the coarse-grained energy dissipation, a slight disadvantage related
to exponential moments is identified. We conjecture the existence of a
more appropriate infinitely divisible distribution without this
disadvantage.

%
%
The energy dissipation exhibits stylised features beyond the scaling
and self-scaling of the two-point correlators.
In~\cite{Cl-Sc-Gr-2008}, breakdown coefficients and Kramers-Moyal
coefficients are employed, in particular, to evaluate the use of
stable distributions in the modelling. A similar analysis remains to
be performed in the case of the normal inverse Gaussian distributions.


\appendix
\section{The normal inverse Gaussian distribution}
\label{app:nig}

%
%
The normal inverse Gaussian distributions form a four-parameter family
of probability distributions on the real line. They are a special case
of the generalised hyperbolic distributions introduced
in~\cite{BaNi-1977} to describe the law of the logarithm of the size
of sand particles (see
also~\cite{BaNi-Ha-1977,BaNi-1978,BaNi-Bl-1981,Bl-1990,Eb-Ha-2004}).
The generalised hyperbolic distributions are applied in many areas of
science, see e.g.~\cite{Pr-1999,BaNi-Bl-Sc-2004} and the references
therein.

%
%
The probability density function of a normal inverse Gaussian
distribution is given by
\begin{align}
  \label{eq:NIGpdf}
  \text{pdf}_{\NIG(\alpha,\beta,\mu,\delta)}(x)
  &= \frac{\alpha e^{\delta\gamma}}{\pi} e^{\beta(x-\mu)}
  \frac{K_1\bigl(\delta\alpha q\bigl(\frac{x-\mu}{\delta}\bigr)\bigr)}
  {q\bigl(\frac{x-\mu}{\delta}\bigr)}
\end{align}
where $\gamma=\alpha^2-\beta^2$, $q(x)=\sqrt{1+x^2}$, and $K_1$
denotes the modified Bessel function of the second kind with
index~$1$. The real parameter~$\mu$ determines the location, and the
positive parameter~$\delta$ determines the scale. The
parameters~$\alpha$ and~$\beta$ are shape parameters and lie within
the shape cone: $|\beta|<\alpha$.

%
%
From the asymptotic property $K_1(x)\propto x^{-1/2}e^{-x}$ as
$x\to\infty$ it follows that the NIG distribution has semi-heavy
tails, specifically
\begin{align}
  \label{eq:semi-heavy}
  \pdf_{\NIG(\alpha,\beta,\mu,\delta)}(x)
  \propto |x|^{-3/2}\exp(-\alpha|x|+\beta x)
\end{align}
as $x\to\pm\infty$. This illuminates the role of~$\alpha$ and~$\beta$
in determining the tails of the distribution.

%
%
The cumulant function $K(t;\alpha,\beta,\mu,\delta) = \log\<\exp(t X)\>$
of a random variable~$X$ with distribution
$\NIG(\alpha,\beta,\mu,\delta)$ is given by
\begin{align}
  \label{eq:NIGK}
  K(t;\alpha,\beta,\mu,\delta)
  = \mu t 
  + \delta\bigl(\gamma-\sqrt{\alpha^2-(\beta+t)^2}\bigr),
\end{align}
and the radius of convergence for the cumulant function is
$|\alpha-\beta|$.

%
%
By differentiating~\eqref{eq:NIGK} we obtain the
following expressions for the first four cumulants,
\begin{align}
  \label{eq:NIG4cum}
  \begin{aligned}
  \kappa_1 
  &= \mu + \delta\frac{\rho}{(1-\rho^2)^{1/2}}, &
  \kappa_2 
  &= \frac{\delta}{\alpha}\frac{1}{(1-\rho^2)^{3/2}},
  \\
  \kappa_3 
  &= 3\frac{\delta}{\alpha^2}\frac{\rho}{(1-\rho^2)^{5/2}}, &
  \kappa_4 
  &= 3\frac{\delta}{\alpha^3}
  \frac{1+4\rho^2}{(1-\rho^2)^{7/2}},
\end{aligned}
\end{align}
where $\rho=\beta/\alpha$. Hence, the standardised third and fourth
cumulants are
\begin{align}
  \label{eq:stdNIGcum}
    \frac{\kappa_3}{\kappa_2^{3/2}}
    &= 3\frac{\rho}{(\delta\alpha(1-\rho^2)^{1/2})^{1/2}},
    &
    \frac{\kappa_4}{\kappa_2^2}
    &= 3\frac{1+4\rho^2}{\delta\alpha(1-\rho^2)^{1/2}}.
\end{align}
Equations~\eqref{eq:NIG4cum} and~\eqref{eq:stdNIGcum} further
illuminate the roles of the four parameters: $\mu$ and $\delta$
determine location and scale, respectively; $\beta$ is related to the
skewness, specifically the tail asymmetry (if $\beta=0$, the
distribution is symmetric); and $\alpha$ is related to the kurtosis.

%
%
It follows immediately from \eqref{eq:NIGK} that if $X_1,\ldots,X_n$
are independent normal inverse Gaussian variables with common
parameters $\alpha$ and $\beta$, but with individual location and
scale parameters~$\mu_i$ and~$\delta_i$ ($i=1,\ldots,n$), then the
distribution of the sum $X_+=X_1+\cdots+X_n$ is
$\NIG(\alpha,\beta,\mu_+,\delta_+)$ where $\mu_+=\mu_1+\cdots+\mu_n$
and $\delta_+=\delta_1+\cdots+\delta_n$. 
Therefore, the normal inverse Gaussian distributions are
\emph{infinitely divisible}, see also~\cite{BaNi-Ha-1977}.

%
%

\begin{figure}
  \centering
  \includegraphics{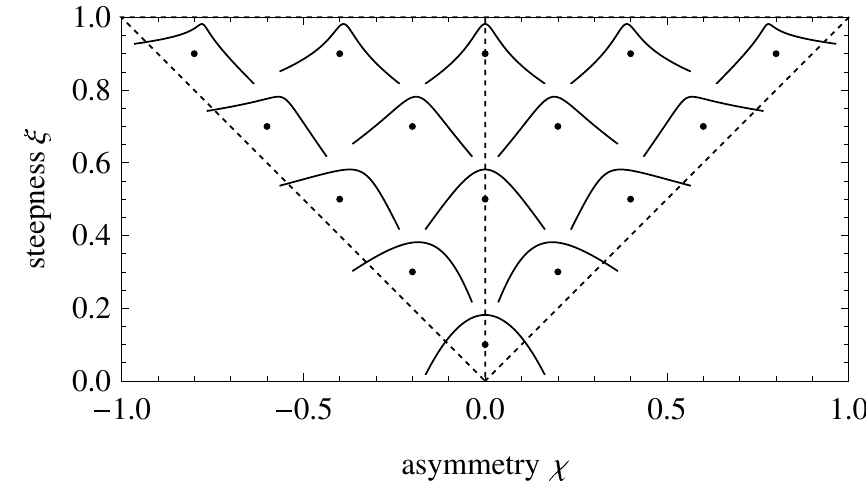}
  \caption{The normal inverse Gaussian shape triangle. The black dots
    mark positions $(\chi,\xi)$ in the shape triangle, and the small
    graph nearby shows the logarithm of the probability density
    function with the corresponding asymmetry~$\chi$, steepness~$\xi$,
    zero mean, and unit variance, plotted on the interval $[-3,3]$.}
  \label{fig:nig-tri}
\end{figure}

It is often desirable to describe the NIG distributions in terms of
location-scale invariant parameters. By letting
$\bar\alpha=\delta\alpha$ and $\bar\beta=\delta\beta$ we have
that~$\bar\alpha$ and~$\bar\beta$ are invariant under change of
location and scale. For the purpose of interpreting the shape
parameters it is sometimes advantageous to express the shape in terms
of the location-scale invariant \emph{steepness~$\xi$} and
\emph{asymmetry~$\chi$}, defined by
\begin{align}
  \xi 
  &= \frac1{\sqrt{1+\bar\gamma}},
  &
  \chi
  &=\rho\xi=\frac\rho{\sqrt{1+\bar\gamma}},
\end{align}
where $\bar\gamma=\delta\gamma=\delta\sqrt{\alpha^2+\beta^2}$, and
$\rho=\beta/\alpha=\bar\beta/\bar\alpha$ is the \emph{alternative
  asymmetry}. These parameters are within the \emph{shape triangle},
defined by
\begin{align}
  \label{eq:NIGshapetri}
  \{(\chi,\xi)\mid 0<\xi<1,-\xi<\chi<\xi\}.
\end{align}
Figure~\ref{fig:nig-tri} shows the shape of the normal inverse
Gaussian distributions for various values of the asymmetry~$\chi$ and
steepness~$\xi$. A wide range of shapes is possible. (See
\cite{Bl-1990} for details on the shape of the family of generalised
hyperbolic distributions).

%
%
If the variance is held constant, the normal distribution is obtained
in the limit of zero steepness. For details and examples of other
limiting distributions, see~\cite{BaNi-1978,Eb-Ha-2004}.
%
%
\begin{figure}
  \centering
  \includegraphics{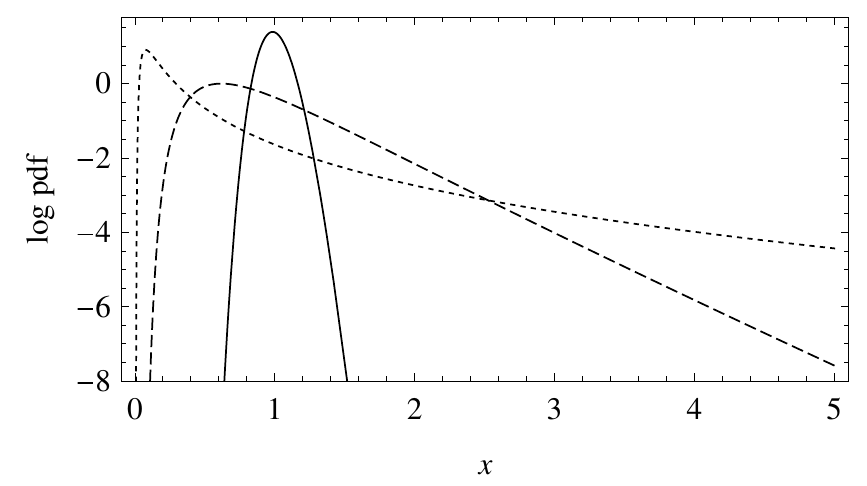}
  \caption{The logarithm of the probability density function of the
    inverse Gaussian distribution with parameters chosen to yield
    normal inverse Gaussian distributions with zero mean, unit
    variance, zero asymmetry, and steepness $\xi=0.1$ (black),
    $\xi=0.5$ (long dashes), and $\xi=0.9$ (short dashes), see
    fig.~\protect\ref{fig:nig-tri}.}
  \label{fig:ig-pdf}
\end{figure}

A useful property of the normal inverse Gaussian distribution is the
representation in terms of a mean-variance mixture of a normal
distribution with the mixing distribution being an inverse Gaussian
distribution, hence the name. Specifically, if $X$ is normal with mean
$\mu+\beta\sigma^2$ and variance $\sigma^2$, and if $\sigma^2$ is
endowed with an independent inverse Gaussian distribution with
parameters $\gamma$ and $\delta$, then $X$ follows a normal inverse
Gaussian distribution with parameters $(\alpha,\beta,\mu,\delta)$. In
short we may write
\begin{align}
  \label{eq:NIGmix}
  X \sim \mu + \beta\sigma^2 + \sigma U
\end{align}
where $U$ is a normal distribution with zero mean and unit variance.
For reference, the probability density function of an inverse Gaussian
distribution with parameters $\gamma$ and $\delta$ is
\begin{align}
  \label{eq:IGpdf}
  \text{pdf}_{\text{IG}(\gamma,\delta)}(x)
  &= \frac{\delta e^{\gamma\delta}}{\sqrt{2\pi}} x^{-3/2}
  \exp(-\tfrac12(x\gamma^2+x^{-1}\delta^2))
\end{align}
for $x>0$. Figure~\ref{fig:ig-pdf} shows the logarithm of the
probability density functions of the inverse Gaussian distribution
corresponding to the three symmetric normal inverse Gaussian
distributions in fig.~\ref{fig:nig-tri}. As the steepness increases,
the probability that the random variance $\sigma^2$
in~\eqref{eq:NIGmix} will attain large values increases.

%
%
To estimate the normal inverse Gaussian parameters from data one may
apply maximum (or pseudo) likelihood methods. The maximum likelihood
estimation is non-trivial since the likelihood function is very flat
near the optimum. The computer program ``hyp''~\cite{Bl-So-1992}
implements numerical maximisation of the likelihood function, but, in
general, non-linear optimisation algorithms may also be applied. The
approach of expectation-maximisation developed in~\cite{Ka-2002} may
be used in conjunction with other optimisation algorithms.


\section{Integration with respect to L\'evy bases}
\label{app:levy}

%
%
The stochastic processes to be considered in the present paper are
expressed in terms of integrals of deterministic functions with
respect to L\'evy bases on~$\R^2$. A L\'evy basis~$Z$ on $\R^2$ is an
\emph{infinitely divisible, independently scattered random measure},
i.e., to each bounded Borel subset~$A$ of $\R^2$, an infinitely
divisible random variable $Z(A)$ is associated, the random variables
associated to disjoint subsets are independent, and the random
variable associated to a disjoint union is almost surely equal to
the sum of the random variables associated to each subset, provided
the union is a bounded Borel subset. For more details and mathematical
rigour we refer to~\cite{Ra-Ro-1989}.

%
%
The stochastic integral $\int f\,dZ$ of a deterministic function
$f\colon\R^2\to\R$ with respect to a L\'evy basis~$Z$ is defined in
three steps. Firstly, for an indicator function $\1_A$ we define $\int
\1_A\,dZ = Z(A)$. Secondly, by requiring that the integral is linear
in the integrand, the integral is extended to simple functions, i.e.,
linear combinations of indicator functions. Finally, since a
measurable function $f\colon\R^2\to\R$ may be approximated by a
sequence of simple functions, the integral $\int f\,dZ$ is defined to
be the limit in probability of the sequence of integrals of the simple
functions, provided this limit exists.

%
%

An important class of L\'evy bases has the property that, informally,
the distribution of the random variable associated to a subset does
not depend on the location of the subset. In this case, we have the
following fundamental representations of the Laplace transform and the
characteristic function of the integral of a deterministic
function~$f$ with respect to a L\'evy basis~$Z$. Let $K(s\ddagger X) =
\log\<\exp(sX)\>$ and $C(s\ddagger X) = \log\<\exp(isX)\>$ denote the
logarithm of the Laplace transform and the characteristic function of
the random variable~$X$, respectively. Then
\begin{align}
  K\Bigl(s\ddagger\int_A f(a)\,Z(da)\Bigr)
  &= \int_A K(sf(a)\ddagger Z')\,da,
  \label{eq:kumulant}
  \\
  C\Bigl(s\ddagger\int_A f(a)\,Z(da)\Bigr)
  &= \int_A C(sf(a)\ddagger Z')\,da,
  \label{eq:cumulant}
\end{align}
where $Z'$ is a random variable (called the \emph{L\'evy seed}) whose
cumulant function is related to the L\'evy basis $Z$ by 
\begin{align}
  K(s\ddagger Z(da)) &= K(s\ddagger Z')\,da,
  \\
  C(s\ddagger Z(da)) &= C(s\ddagger Z')\,da,
\end{align}
see~\cite{BaNi-Sc-2004} for more details. It follows that the
distribution of the stochastic integral is determined by the
function~$f$ and the log-characteristic function of the L\'evy
seed~$Z'$.

\section*{Acknowledgements}

The authors wish to thank B.~Chabaud for granting permission to use
the data. The authors furthermore acknowledge useful comments from Ole
E. Barn\-dorff-Niel\-sen and M. Greiner.


\bibliographystyle{plain}
\bibliography{energy-dissipation}

\begin{thebibliography}{10}

\bibitem{Ar-Ba-Mu-1998}
A.~Arneodo, E.~Bacry, and J.~F. Muzy.
\newblock Random cascades on wavelet dyadic trees.
\newblock {\em J. Math. Phys.}, 39:4142--4164, 1998.

\bibitem{BaNi-1978}
O~E Barndorff-Nielsen.
\newblock Hyperbolic distributions and distributions on hyperbolae.
\newblock {\em Scand. J. Statist.}, 5(3):151--157, 1978.

\bibitem{BaNi-Ha-1977}
O~E Barndorff-Nielsen and Christian Halgreen.
\newblock Infinite divisibility of the hyperbolic and generalized inverse
  {G}aussian distributions.
\newblock {\em Z. Wahrscheinlichkeitstheorie verw. Gebiete}, 38(4):309--311,
  1977.

\bibitem{BaNi-Sc-2004}
O~E Barndorff-Nielsen and J~Schmiegel.
\newblock L{\'e}vy-based spatial-temporal modelling, with applications to
  turbulence.
\newblock {\em Russian Math. Surveys}, 59(1):65, 2004.

\bibitem{BaNi-1977}
Ole~E Barndorff-Nielsen.
\newblock Exponentially decreasing distributions for the logarithm of particle
  size.
\newblock {\em Proc. R. Soc. Lond. A}, 353(1674):401--419, 1977.

\bibitem{BaNi-Bl-1981}
Ole~E Barndorff-Nielsen and P~Bl{\ae}sild.
\newblock Hyperbolic distributions and ramifications: Contributions to theory
  and application.
\newblock In C~Taillie, G~Patil, and B~Baldessari, editors, {\em Statistical
  Distributions in Scientific Work}, volume~4, pages 19--44. Reidel, Dordrecht,
  1981.

\bibitem{BaNi-Bl-Sc-2004}
Ole~E Barndorff-Nielsen, P~Bl{\ae}sild, and J{\"u}rgen Schmiegel.
\newblock {A parsimonious and universal description of turbulent velocity
  increments}.
\newblock {\em Eur. Phys. J. B}, 2004.

\bibitem{BaNi-Ke-Soe-1982}
Ole~E Barndorff-Nielsen, John Kent, and M~S{\o}rensen.
\newblock Normal variance-mean mixtures and {$z$} distributions.
\newblock {\em Int. Stat. Rev.}, 50(2):145--159, 1982.

\bibitem{Be-et-al-1993b}
R~Benzi, L.~Biferale, A.~Crisanti, G.~Paladin, M.~Vergassola, and A.~Vulpiani.
\newblock A random process for the construction of multiaffine fields.
\newblock {\em Physica D}, 65:352--358, 1993.

\bibitem{Be-et-al-1993}
R~Benzi, S~Ciliberto, C~Baudet, G.R. Chavarria, and R~Tripiccione.
\newblock {Extended self-similarity in the dissipation range of fully developed
  turbulence}.
\newblock {\em Europhys. Lett.}, 24(4):275--279, 1993.

\bibitem{Be-Pa-Pa-Vu-1984}
R~Benzi, G~Paladin, G~Parisi, and A~Vulpiani.
\newblock On the multifractal nature of fully developed turbulence and chaotic
  systems.
\newblock {\em J. Phys. A}, 17(18):3521, 1984.

\bibitem{Bi-et-al-1998}
L.~Biferale, G.~Boffetta, A.~Celani, A.~Crisanti, and A.~Vulpiani.
\newblock Mimicking a turbulent signal: Sequential multiaffine processes.
\newblock {\em Phys. Rev. E}, 57(6):R6261--R6264, 1998.

\bibitem{Bl-1990}
P~Bl{\ae}sild.
\newblock {The shape cone of the d-dimensional hyperbolic distribution}.
\newblock Technical report, Dep. of Theoretical Statistics, Inst. of
  Mathematics, Univ. of Aarhus, 1990.

\bibitem{Bl-So-1992}
P~Bl{\ae}sild and M~K S{\o}rensen.
\newblock {Hyp: A Computer Program for Analyzing Data by Means of the
  Hyperbolic Distribution}.
\newblock Technical report, Dep. of Theoretical Statistics, Inst. of
  Mathematics, Univ. of Aarhus, 1992.

\bibitem{Ch-et-al-2000}
O.~Chanal, B.~Chabaud, B~Castaing, and B.~H{\'e}bral.
\newblock {Intermittency in a turbulent low temperature gaseous helium jet}.
\newblock {\em Eur. Phys. J. B}, 17(2):309--317, 2000.

\bibitem{Cl-Gr-Sr-2003}
J.~Cleve, M.~Greiner, and K.~R. Sreenivasan.
\newblock On the effects of surrogacy of energy dissipation in determining the
  intermittency exponent in fully developed turbulence.
\newblock {\em Europhys. Lett.}, 61(6):756, 2003.

\bibitem{Cl-et-al-2005}
Jochen Cleve, Thomas Dziekan, J{\"u}rgen Schmiegel, Ole~E Barndorff-Nielsen,
  Bruce~R Pearson, Katepalli~R Sreenivasan, and Martin Greiner.
\newblock {Finite-size scaling of two-point statistics and the turbulent energy
  cascade generators}.
\newblock {\em Phys. Rev. E}, 71(2):1--12, 2005.

\bibitem{Cl-Gr-2000}
Jochen Cleve and Martin Greiner.
\newblock The markovian metamorphosis of a simple turbulent cascade model.
\newblock {\em Phys. Lett. A}, 273:104 -- 108, 2000.

\bibitem{Cl-et-al-2004}
Jochen Cleve, Martin Greiner, Bruce~R Pearson, and Katepalli~R Sreenivasan.
\newblock {Intermittency exponent of the turbulent energy cascade}.
\newblock {\em Phys. Rev. E}, 69(6):1--6, 2004.

\bibitem{Cl-Sc-Gr-2008}
Jochen Cleve, J{\"u}rgen Schmiegel, and Martin Greiner.
\newblock Apparent scale correlations in a random multifractal process.
\newblock {\em Eur. Phys. J. B}, 63(1):109--116, 2008.

\bibitem{Du-1994}
B{\'e}reng{\`e}re Dubrulle.
\newblock {Intermittency in fully developed turbulence: Log-Poisson statistics
  and generalized scale covariance}.
\newblock {\em Phys. Rev. Lett.}, 73(7):959--962, 1994.

\bibitem{Eb-Ha-2004}
Ernst Eberlein and Ernst August~v. Hammerstein.
\newblock Generalized hyperbolic and inverse gaussian distributions: limiting
  cases and approximation of processes.
\newblock In R.~C. Dalang, M.~Dozzi, and F.~Russo, editors, {\em Seminar on
  Stochastic Analysis, Random Fields and Applications IV}, volume~58 of {\em
  Progress in Probability}, pages 221--264. {Birkh\"auser Verlag}, 2004.

\bibitem{El-et-al-1995}
F.W. Elliot, Jr., A.J. Majda, D.J. Horntrop, and R.M McLaughlin.
\newblock {Hierarchical Monte Carlo Methods for Fractal Random Fields}.
\newblock {\em J. Stat. Phys.}, 81(3/4):717--736, 1995.

\bibitem{Fr-1995}
U~Frisch.
\newblock {\em Turbulence}.
\newblock Cambridge Univ. Press, Cambridge, 1995.

\bibitem{Fr-Su-Ne-1978}
Uriel Frisch, Pierre-Louis Sulem, and Mark Nelkin.
\newblock A simple dynamical model of intermittent fully developed turbulence.
\newblock {\em J. Fluid Mech.}, 87(04):719--736, 1978.

\bibitem{Jo-Gr-Li-2000}
B.~Jouault, M.~Greiner, and P.~Lipa.
\newblock Fix-point multiplier distributions in discrete turbulent cascade
  models.
\newblock {\em Physica D}, 136:125 -- 144, 2000.

\bibitem{Jo-Li-Gr-1999}
Bruno Jouault, Peter Lipa, and Martin Greiner.
\newblock Multiplier phenomenology in random multiplicative cascade processes.
\newblock {\em Phys. Rev. E}, 59:2451--2454, 1999.

\bibitem{Ju-et-al-1994}
A.~Juneja, D.P. Lathrop, K.~R. Sreenivasan, and G.~Stolovitzky.
\newblock Synthetic turbulence.
\newblock {\em Phys. Rev. E}, 49(6):5179--5194, 1994.

\bibitem{Ka-2002}
Dimitris Karlis.
\newblock An {EM} type algorithm for maximum likelihood estimation of the
  normal-inverse {G}aussian distribution.
\newblock {\em Statist. Probab. Lett.}, 57(1):43--52, 2002.

\bibitem{Ko-1962}
A~N Kolmogorov.
\newblock {A refinement of previous hypotheses concerning the local structure
  of turbulence in a viscous incompressible fluid at high Reynolds number}.
\newblock {\em J. Fluid Mech}, 13(1):82--85, 1962.

\bibitem{Ma-1974}
Benoit~B. Mandelbrot.
\newblock Intermittent turbulence in self-similar cascades: divergence of high
  moments and dimension of the carrier.
\newblock {\em J. Fluid Mech.}, 62(02):331--358, 1974.

\bibitem{Me-Sr-1991}
Charles Meneveau and K.~R. Sreenivasan.
\newblock The multifractal nature of turbulent energy dissipation.
\newblock {\em J. Fluid Mech.}, 224:429--484, 1991.

\bibitem{Ob-1962}
A~M Oboukhov.
\newblock {Some specific features of atmospheric tubulence}.
\newblock {\em J. Fluid Mech.}, 13(01):77--81, 1962.

\bibitem{Pe-Wa-1993}
Donald~B Percival and Andrew~T Walden.
\newblock {\em Spectral Analysis for Physical Applications: Multitaper and
  Conventional Univariate Techniques}.
\newblock Cambridge University Press, 1993.

\bibitem{Pr-1999}
K~Prause.
\newblock {\em {The Generalized Hyperbolic Model: Estimation, Financial
  Derivatives and Risk Measures}}.
\newblock PhD thesis, Albert-Ludwigs University, 1999.

\bibitem{Ra-Ro-1989}
Balram~S. Rajput and Jan Rosi{\'n}ski.
\newblock Spectral representations of infinitely divisible processes.
\newblock {\em Probab. Th. Rel. Fields}, 82(3):451--487, 1989.

\bibitem{Sa-Ta-1994}
Gennady Samorodnitsky and Murad~S. Taqqu.
\newblock {\em Stable non-Gaussian random processes}.
\newblock Chapman \& Hall, 1994.

\bibitem{Sc-Lo-1987}
D~Schertzer and S~Lovejoy.
\newblock {Physical modeling and analysis of rain and clouds by anisotropic
  scaling multiplicative processes}.
\newblock {\em J. Geophys. Res}, 92(D8):9693--9714, 1987.

\bibitem{Sc-2005}
J{\"u}rgen Schmiegel.
\newblock {Self-scaling of turbulent energy dissipation correlators}.
\newblock {\em Phys. Lett. A}, 337(4-6):342--353, 2005.

\bibitem{Sc-BaNi-Eg-2005}
J{\"u}rgen Schmiegel, Ole~E Barndorff-Nielsen, and H~Eggers.
\newblock {A class of spatio-temporal and causal stochastic processes, with
  application to multiscaling and multifractality}.
\newblock {\em S. Afr. J. Sci.}, 101:513--519, 2005.

\bibitem{Sc-et-al-2004}
J\"urgen Schmiegel, Jochen Cleve, Hans~C. Eggers, Bruce~R. Pearson, and Martin
  Greiner.
\newblock Stochastic energy-cascade model for $(1+1)$-dimensional fully
  developed turbulence.
\newblock {\em Phys. Lett. A}, 320(4):247--253, 2004.

\bibitem{Sh-Le-1994}
Zhen-Su She and Emmanuel Leveque.
\newblock {Universal scaling laws in fully developed turbulence}.
\newblock {\em Phys. Rev. Lett.}, 72(3):336--339, 1994.

\bibitem{Sh-Wa-1995}
Zhen-Su She and Edward Waymire.
\newblock {Quantized Energy Cascade and Log-Poisson Statistics in Fully
  Developed Turbulence}.
\newblock {\em Phys. Rev. Lett.}, 74(2):262--265, 1995.

\bibitem{Sr-An-1997}
K.~R. Sreenivasan and R.~Antonia.
\newblock {The phenomenology of small-scale turbulence}.
\newblock {\em Annu. Rev. Fluid Mech.}, 29:435--472, 1997.

\bibitem{Vi-Ba-1991}
T.~Vicsek and A.-L. Barab{\'a}si.
\newblock Multi-affine model for the velocity distribution in fully developed
  turbulent flows.
\newblock {\em J. Phys. A}, 24:L845--L851, 1991.

\end{thebibliography}

\end{document}